\documentclass[iop,revtex4]{emulateapj}

\usepackage{natbib, hyperref}
\usepackage{amssymb, amsmath}
\usepackage{color}
\usepackage{times}

\long\def\symbolfootnote[#1]#2{\begingroup%
\def\thefootnote{\fnsymbol{footnote}}\footnote[#1]{#2}\endgroup}

\slugcomment{To appear in The Astrophysical Journal}
\shorttitle{Blue Straggler Star Populations in Globular Clusters: I. Dynamical Properties }
\shortauthors{Simunovic \& Puzia}

\begin{document}

\title{Blue Straggler Star Populations in Globular Clusters: I. Dynamical Properties of Blue Straggler Stars in NGC 3201, NGC 6218 and $\omega$ Centauri}
\author{Mirko Simunovic and Thomas H. Puzia}
\affil{Institute of Astrophysics, Pontificia Universidad Cat\'olica de Chile, Av. Vicu\~na Mackenna 4860, 7820436 Macul, Santiago, Chile}
\email{Email: msimunov@astro.puc.cl, tpuzia@astro.puc.cl}






\begin{abstract}
We present the first dynamical study of Blue Straggler Stars (BSSs) in three Galactic globular clusters, NGC\,3201, NGC\,5139 ($\omega$Cen), and NGC\,6218, based on medium-resolution spectroscopy ($R\!\approx\!10000$) obtained with the Inamori-Magellan Areal Camera \& Spectrograph (IMACS) mounted at the 6.5 meter Baade Magellan telescope. Our BSS candidate selection technique uses HST/ACS and ESO/WFI photometric data out to $>\!4.5\,r_c$. We use radial velocity measurements to discard non-members and achieve a success rate of $\sim93\%$, which yields a sample of 116 confirmed BSSs. Using the penalized pixel fitting method (pPXF) we measure the $v\sin(i)$ values of the sample BSSs and find their distribution functions peaked at slow velocities with a long tail towards fast velocities in each globular cluster. About $90\%$ of the BSS population in NGC\,3201 and NGC\,6218 exhibit values in the range $10\!-\!50$ km s$^{-1}$, while about $80\%$ of the BSSs in $\omega$Cen show $v\sin(i)$ values between 20 and 70 km s$^{-1}$. We find that the BSSs in NGC\,3201 and NGC\,6218 which show $v\sin(i)\!>\!50$ km s$^{-1}$ are all found in the central cluster regions, inside a projected $2\,r_c$, of their parent clusters. We find a similar result in $\omega$Cen for BSSs with $v\sin(i)\!>\!70$ km s$^{-1}$ which are all, except for two, concentrated inside $2\,r_c$. In all globular clusters we find rapidly rotating BSSs that have relatively high differential radial velocities which likely put them on hyperbolic orbits, suggestive of strong dynamical interactions in the past. Based on stellar spin down and dynamical crossing timescales we estimate that all the observed rapidly rotating BSSs are likely to form in their central cluster regions no longer than $\sim\!300$ Myr ago and may be subsequently ejected from their host globular clusters. Using dereddened $V\!-\!I$ colors of our photometric selection we show that blue BSSs in $\omega$Cen with $(V\!-\!I)_0\la\!0.25$ mag show a significantly increased $v\sin(i)$ dispersion compared with their red counterparts and all other BSSs in our sample, therefore strongly implying that fast rotating BSSs in $\omega$Cen are preferentially bluer, i.e.~more massive. This may indicate that this particular blue BSS population was formed in an unique formation event and/or through an unique mechanism. 
\end{abstract}

\keywords{globular clusters: general --- globular clusters: individual (NGC\,3201, NGC\,5139, NGC\,6218)}



\section{Introduction}

\subsection{The Nature of Blue Straggler Stars}
Globular Clusters (GCs) are excellent laboratories of stellar evolution and host abundant representatives of some of the more exotic stellar evolutionary phases.~One such, still relatively unexplored phase is known by the name of its representatives, namely Blue Straggler Stars (BSSs).~BSSs are characterized by their effective temperatures and luminosities as members of the main-sequence stellar population.~However, given the ages of their host GCs they are expected to have evolved off the main-sequence to become red giant stars long time ago as their Hertzsprung-Russell diagram parameters place them at higher temperatures and luminosities, thus higher stellar masses, with respect to the main-sequence turn-off point \citep{stryker93, bailyn95}.~The existence of BSSs implies that they must form in more recent events, after the majority of the constituent GC stellar population was formed. 

\subsection{BSS Formation Mechanisms}
In the recent literature, two formation mechanisms have gained most acceptance and are currently lively debated: $i)$ BSSs form in stellar mergers induced by stellar collisions or binary interaction \citep{lombardi02}, or $ii)$ BSSs are rejuvenated stars forming by mass transfer in a binary system \citep[see][for a detailed review]{sills10}. Both of these scenarios are believed to be actively at work and their predominance being a function of the local environment.~It was first suggested that collision-induced BSSs are expected to form in the high density parts of GCs while mass-transfer BSSs are thought to form in the loose outskirts of GCs \citep{fusi92}.~Although first efforts aimed at measuring which of the two mechanisms dominated \citep{knigge09}, there is recent evidence that both mechanisms are simultaneously at work in GCs \citep{leigh11}. Recently, \cite{ferraro09} confirmed the presence of two distinct BSS sequences in M30 in a deep color-magnitude diagram (CMD) study, which the authors believe is suggestive of different formation histories/mechanisms for each BSS sequence. M30 is thought to have undergone core-collapse which could be responsible for the observed two BSS sequences in the CMD. In this respect, being able to distinguish between different BSS formation processes could become a powerful tool to constrain the GC dynamical evolution history \citep[see also][]{ferraro12}. 

\subsection{BSS Formation Diagnostics}
Previous attempts at detecting a certain BSS formation mechanism is the C-O depletion measured in six 47\,Tuc BSSs \citep{ferraro06a}. Such C-O depleted BSSs are expected if during their formation they accreted CNO-processed material from their binary companions.~Therefore, such chemical anomalies may directly point to a mass-transfer mechanism. However, most of the BSSs in the sample were located in the outer regions of 47\,Tuc, where mass-transfer induced BSS formation is expected to be the dominant mechanism, and still only 14\% of the studied BSSs were found having this C-O depletion (i.e.~mass-transfer) signature.~It is clear that the problem of BSS formation is far from being fully understood, and it is therefore fundamental to perform similar chemical analyses, including other elements, of BSSs in larger samples of GCs, significantly increasing the sample statistics. \\

Another potential indicator of the BSS formation mechanism may be their rotational velocity distribution function. In the mass-transfer scenario, the resulting BSS is believed to conserve most of the angular momentum from the binary system \citep{sarna96}, therefore becoming a fast rotator ($v_{\rm rot}\!>\!50$ km s$^{-1}$), as well as in the collision-induced formation scenario, in which it is also claimed that BSSs are formed having high rotational velocities \citep{benz87}. However, models show that in both cases BSSs could lose most of their initial angular momentum through mechanisms such as accretion disk braking/locking \citep{sills05}. This means that being able to link certain rotational velocity distributions to a certain formation mechanism is rather still complex.~Previous surveys of rotational velocities of BSSs include \cite{lovisi10} who found that 40\% of the measured BSSs in M4 are fast rotators, which is the largest frequency of rapidly rotating BSSs ever detected in a GC, although no C-O depletion was found in any of these BSSs.~Another study \citep{ferraro06b} shows a flat radial distribution for the BSSs in NGC\,5139 ($\omega$Cen), suggesting that this cluster is not completely relaxed yet, and therefore its BSS population would have been mainly formed in a mass-transfer scenario in binaries.~The same group has completed a spectroscopic survey of 78 BSSs in $\omega$Cen and found that the majority of them have rotational velocities $v\sin(i)\!<\!20$ km s$^{-1}$, but still a considerable fraction (30\%) are fast rotators of which some rotate even faster than 100 km s$^{-1}$ \citep{lovisi13a}. Following the intriguing BSS double-sequence in M30 mentioned earlier, 12 BSSs were also observed in this cluster and most of them were found to rotate slowly \citep{lovisi13b}.~Yet another example is the rotational velocity distribution measured for NGC\,6397 \citep{lovisi12} which shows most BSSs to rotate very slowly, and only one of them being a fast rotator. The apparent tendency is therefore a similar BSS rotational velocity distribution function for many GCs, usually peaked around 10 km s$^{-1}$, and a low fraction of fast rotators, except for the cases of M4 and $\omega$Cen. The clear link between these observations and a definitive formation mechanism is still very unclear, since neither chemical analysis nor models have been able to successfully interpret these results. This rather complex scenario calls for larger surveys simultaneously sampling dynamical and chemical properties of BSSs with a homogeneous dataset in order to get a wider picture of the BSS formation processes and to improve on the statistical significance of previous results.~In that context, we carry out this study which presents the first results of a pilot program intended to set constraints on the formation mechanisms of BSSs in GCs based on their spectral characteristics.~We select three GCs:~\objectname{NGC\,3201} ($\alpha_{2000}\!=\!10$h17m36.82s, $\delta_{2000}\!=\!-46$d24\arcmin44.9\arcsec), \objectname{NGC\,6218} ($\alpha_{2000}\!=\!$16h47m14.18s, $\delta_{2000}\!=\!-01$d56\arcmin54.7\arcsec),~and~\objectname{NGC\,5139} ($\omega$Cen; $\alpha_{2000}$=13h26m47.24s, $\delta_{2000}\!=\!-47$d28\arcmin46.5\arcsec).~Their BSS populations have not been previously characterized spectroscopically, except for the case of $\omega$Cen as described above.  

This paper is organized as follows: Section 2 describes the observations and data reduction process. Section 3 analyzes the radial and rotational velocities of BSSs. In Section 4 we discuss the properties of BSSs in the context of our results, and summarize the main result and conclusions in Section 5.

\begin{figure*}[!h]
\centering
\includegraphics[width=7.5cm]{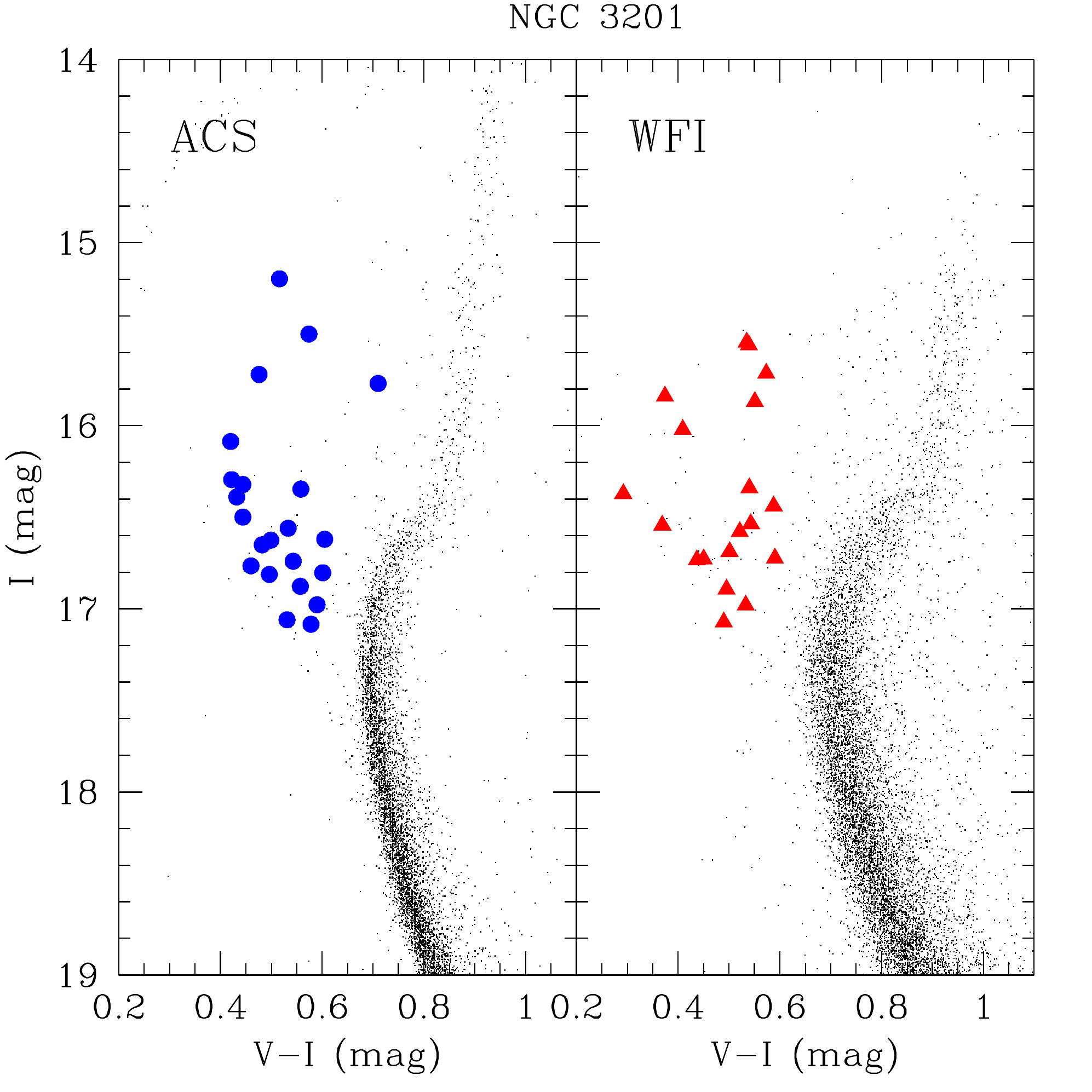}
\includegraphics[width=7.5cm]{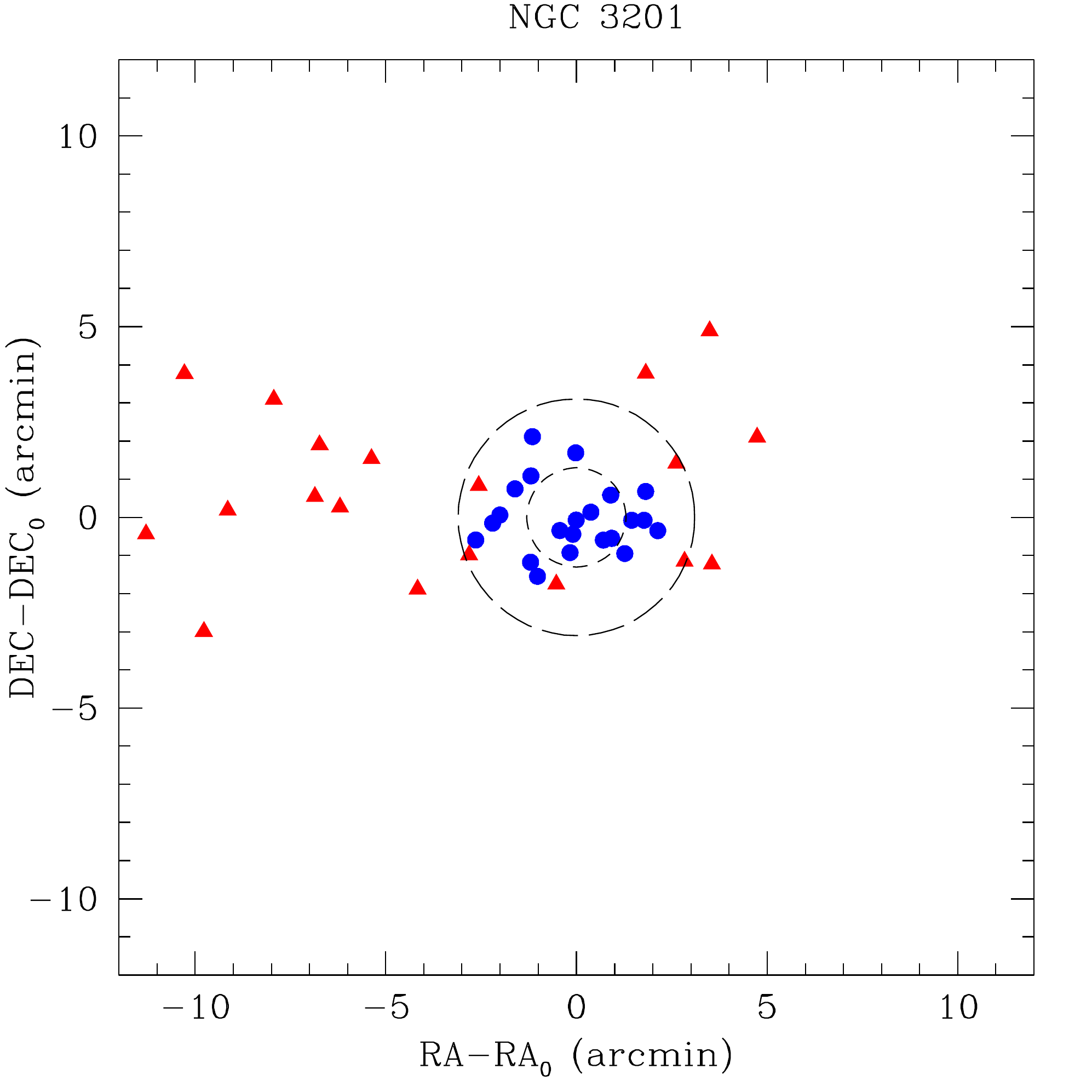}
\includegraphics[width=7.5cm]{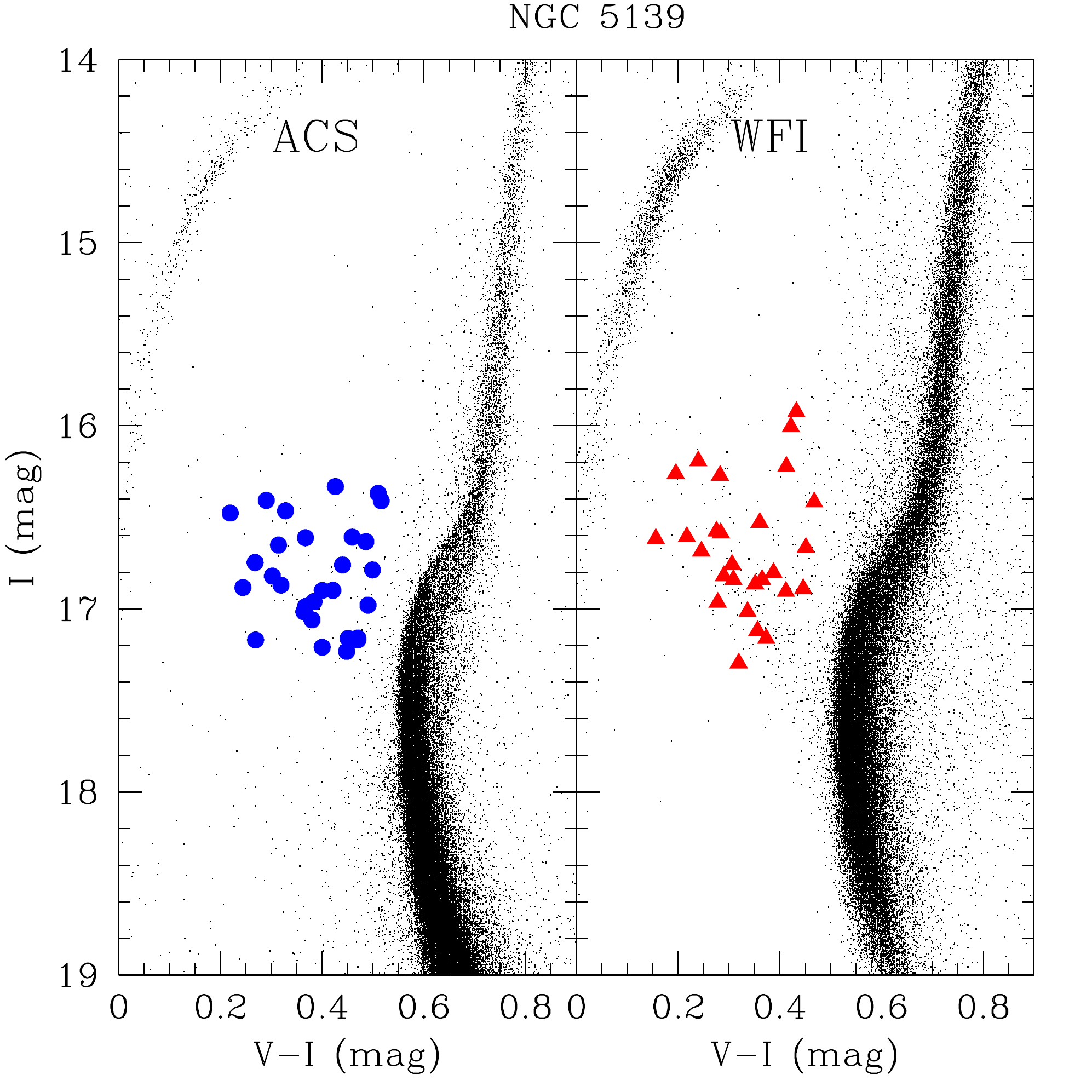}
\includegraphics[width=7.5cm]{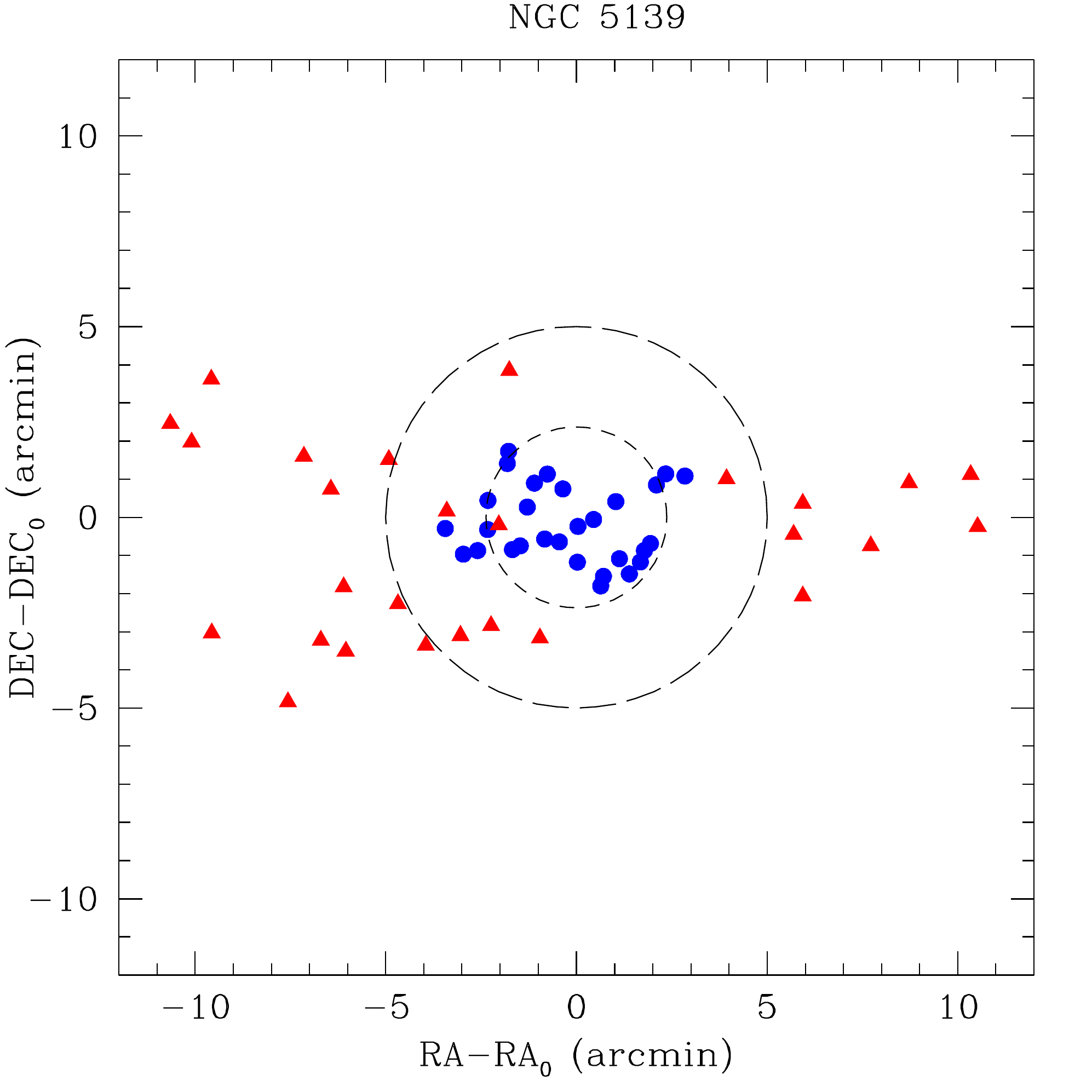}
\includegraphics[width=7.5cm]{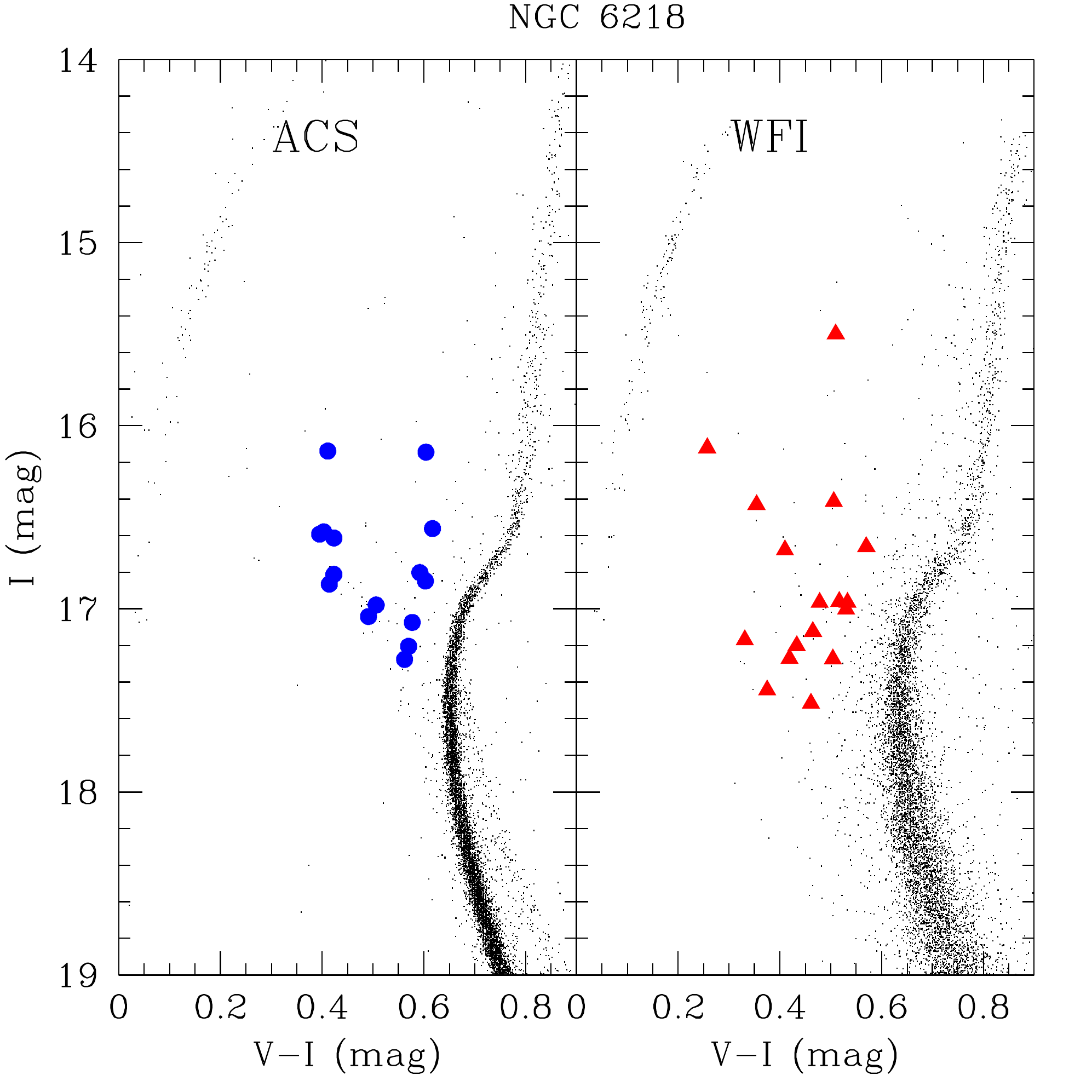}
\includegraphics[width=7.5cm]{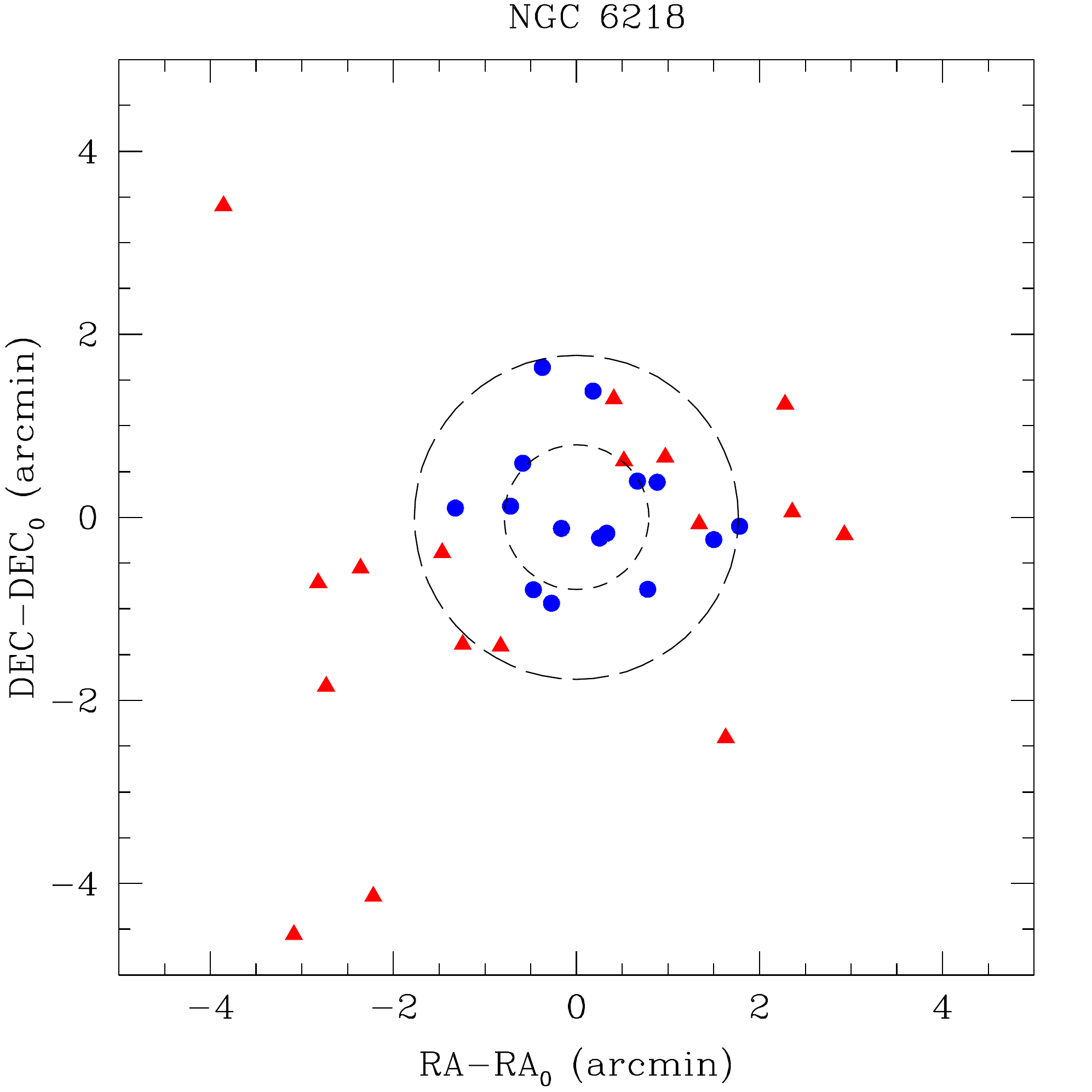}
\caption{({\it Left column}): Color-magnitude diagrams (CMDs) of our target clusters NGC\,3201 ({\it top row}), NGC\,5139 ({\it middle row}), and NGC\,6218 ({\it bottom row}) obtained from ACS photometry ({\it left sub-panels}) and WFI photometry ({\it right sub-panels}), with the selection of BSS candidates labeled as circles and triangles for the ACS ({\it blue circles}) and WFI data ({\it red triangles}), respectively. ({\it Right column}): Spatial locations of the BSSs in each GC with respect to the cluster center position. Inner and outer dashed circles correspond to the cluster core radius and half-light radius, respectively, as reported in \cite{harris96}.}
\label{cmd1}
\end{figure*}

\begin{figure*}[ht!]
\centering
\includegraphics[width=18cm]{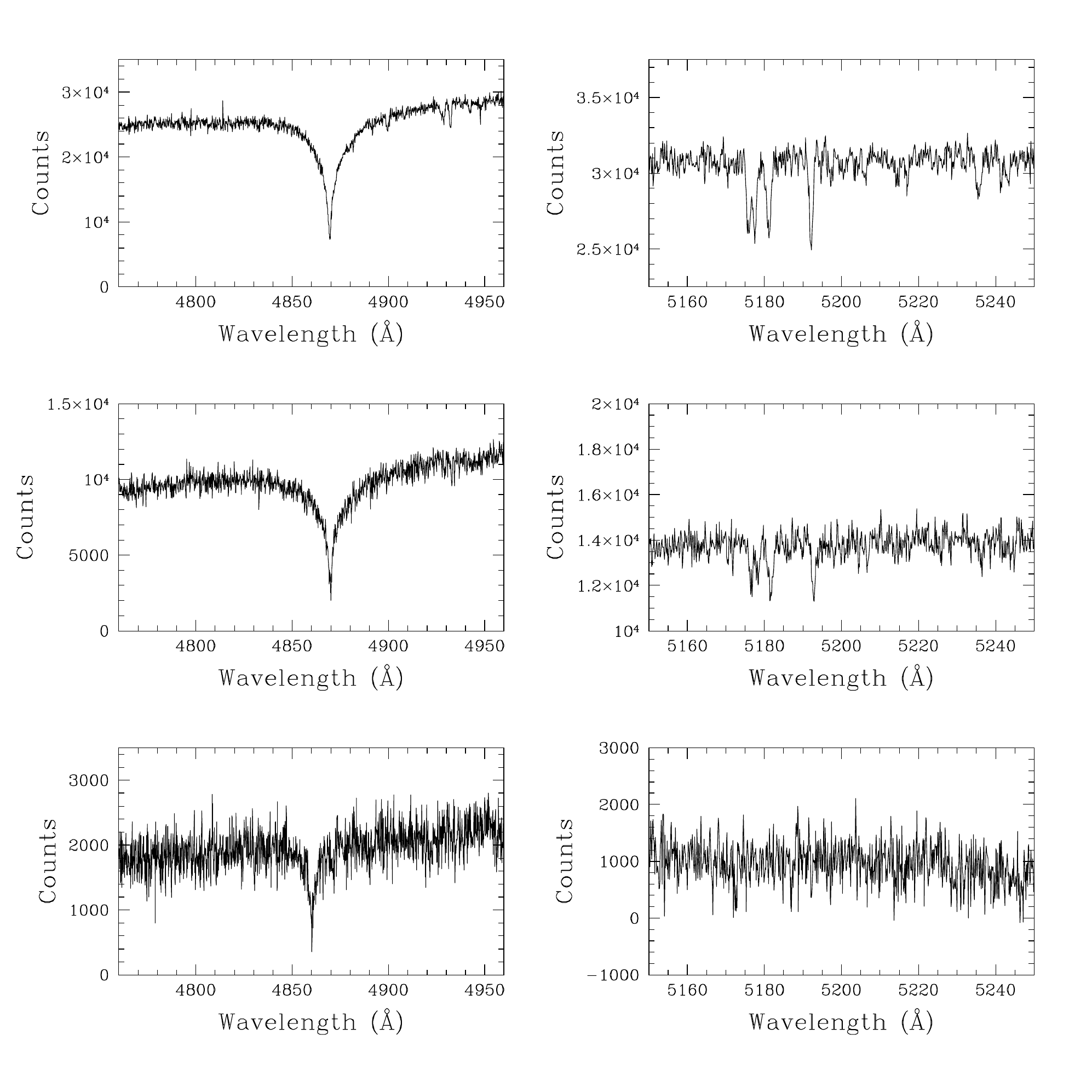}
\caption{Example spectra of BSSs for spectral ranges around H${\beta}$ ({\it left column}) and the Mg$b$ triplet ({\it right column}).~We show three representative cases corresponding to approximately the maximum, the median, and the minimum values of the S/N distribution, illustrated in Figure~\ref{SNR_dist}. The top panels correspond to a S/N$\approx$160, the middle panels are for a S/N$\approx$70 and the bottom panels correspond to a S/N$\approx$20.}
\label{example_spectra}
\end{figure*}

\section{Observations and Data Reduction}

\subsection{Instrumental Setup and Target Selection}
The data used in this work were obtained as part of the observing program CNTAC2012A-101 using the multi-slit mode of the IMACS f/4 spectrograph mounted on the 6.5-meter Baade Telescope at Las Campanas Observatory.~The observations were performed on the night of March 21st, 2012, with overall good sky conditions and an average seeing of 0.6\arcsec.~Multi-slit masks were cut to obtain spectra of 42, 61 and 34 BSS candidates in NGC\,3201, $\omega$Cen and NGC\,6218, respectively. The slit mask configurations were designed such that the center of each GC was placed near the slit-mask center in order to homogeneously obtain spectra of BSS candidates located at as much as $\sim\!10\arcmin$ of radial cluster-centric distance.~In this manner we radially sampled the targets up to approximately 8.5, 4.6, and 6.8\,$r_c$ in NGC\,3201, $\omega$Cen, and NGC\,6218, respectively.~The science observations consisted of 20-minute sub-exposures, from which we obtained total exposure times of 1h:40m for NGC\,3201 and NGC\,6218, and 1h:20m for $\omega$Cen, providing a mean signal-to-noise ratio of S/N~$\!\approx\!70$.~Interspersed between the science target observations, we also obtained spectra for two photometric standards, five rotational velocity standards and arc calibration frames. Standard bias and flat-field frames were obtained at the beginning and at the end of the night.~The instrumental spectrograph configuration for all multi-object science exposures consisted of 4\arcsec\ long and 0.7\arcsec\ wide slits and the 1200 l/mm grism blazed at 17.45$^{\rm o}$, which yields a wavelength coverage of $\sim$3600-5200\,\AA\ with a spectral resolution of $R\!\approx\!10000$ and a resulting velocity resolution of $\sim\!15$ km/s. The corresponding instrumental dispersion relation is $\sim\!0.2$ \AA/pix.
\\

The spectroscopic BSS target selection was performed on F606W ($\sim$$V$) and F814W ($\sim$$I$) band photometric catalogs from the {\it HST/ACS Galactic GC Survey} \citep{sarajedini07} combined with catalogs generated using wide-field imager (WFI) data obtained from the ESO science archive\footnote{http://archive.eso.org} plus a large WFI photometric catalog of $\omega$Cen available online \citep{bellini09}.~The reason for combining HST and WFI catalogs is that the wide field channel (WFC) of the ACS camera samples only the inner $\sim3\arcmin\times3\arcmin$ of the target GCs while the WFI data extends well outside the $\sim15\arcmin\times15\arcmin$ field of view of the IMACS f/4 camera.~By combining these two datasets we guarantee homogeneous sampling of the dense core regions and the outskirts of each GC. The BSS candidates were selected from making color and magnitude cuts in the CMDs of each GC. The selection criteria is the one from \cite{leigh09} which uses the $V$ and $I$ filter magnitudes to determine regions in the CMD that are exclusive to specific stellar evolutionary phases.~These selection criteria and the physical limits of the mask design, set by the slit length and width, resulted in a final selection of 137 BSS candidates.~Figure~\ref{cmd1} shows the CMDs of all three GCs and the spatial distribution of all photometrically selected BSS candidates using HST/ACS and ESO/WFI data.

\begin{figure}[ht!]
\centering
\includegraphics[width=8.9cm]{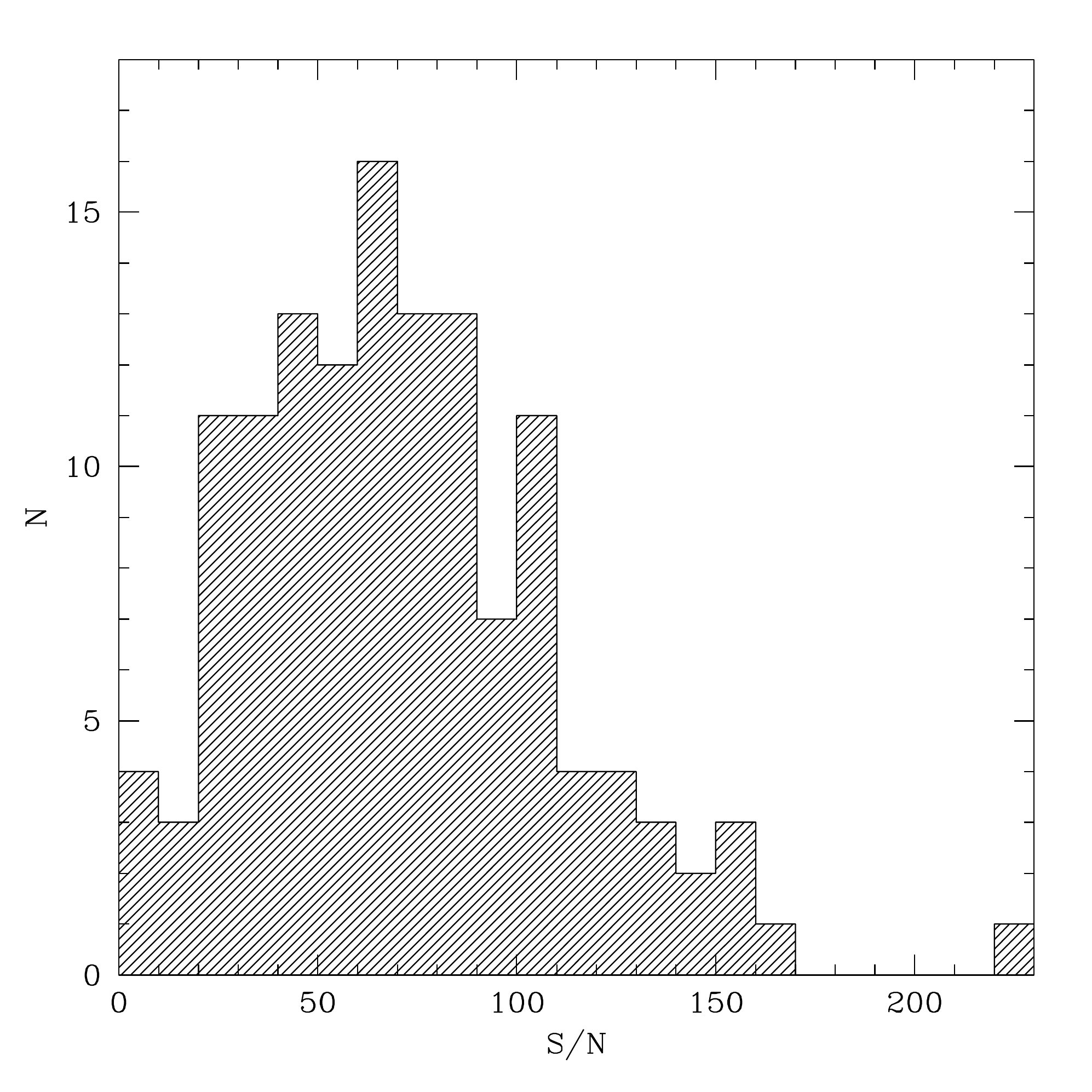}
\caption{The distribution of the median signal-to-noise (S/N) values of all our sample BSS candidate spectra.~Approximately half of our spectra have S/N values larger than $\sim\!70$ and 20\% of them have S/N~$>100$.}
\label{SNR_dist}
\end{figure}

\subsection{Data Reduction}
The reduction of the data was performed in different levels in order to obtain 1-dimensional spectra for each individual BSS candidate from the original frames that contain the multi-slit exposures.~The major part of this work was performed using the {\sc COSMOS} reduction software (v2.18)\footnote{The routines were obtained through the Carnegie Observatory Software Repository, located at http://code.obs.carnegiescience.edu/cosmos.}, which is a suite of programs dedicated for the reduction of multi-slit spectroscopy data obtained with the IMACS instrument on the Baade Magellan Telescope.~Standard bias and flat-field calibration is performed with {\sc COSMOS} sub-routines.

This package allows for an accurate prediction of the location of spectral features of individual slits thanks to a series of alignment and calibration procedures based on raytracing models of the instrument. These models provide a map that is accurate down to $\sim\!2\!-\!3$ pixels.~The program then uses the arc frames to correct these predictions to an accuracy of a fraction of a pixel, constructing the final field distortion map that is then used for the science data wavelength calibration. The final wavelength solution we obtain is accurate to about $\sim\!0.2\!-\!0.3$ pixels for all targets.~We then apply the corresponding calibrations and subtract the sky through different tasks in {\sc COSMOS} and get the individual 2-dimensional spectra for each target. These spectra are then resampled to a linear wavelength scale with a dispersion of 0.1 \AA/pix. The final 1-dimensional extraction is then performed using IRAF\footnote{IRAF is distributed by the National Optical Astronomy Observatories, which is operated by the Association of Universities for Research in Astronomy, Inc. (AURA) under cooperative agreement with the National Science Foundation} task {\sc apall}. We show in Figure~\ref{example_spectra} the H${\beta}$ and Mg$b$ triplet spectral regions for different BSSs spanning the representatives values of S/N, as obtained from its distribution shown in Figure~\ref{SNR_dist}.~We discard spectra with S/N$<$20 and others affected by contamination, resulting in sample sizes of 41, 49 and 32 BSS candidates in NGC 3201, $\omega$Cen and NGC 6218, respectively.

\section{Analysis}

\subsection{Radial Velocities and Cluster Membership}
\begin{figure}[t!]
\centering
\includegraphics[width=8.9cm]{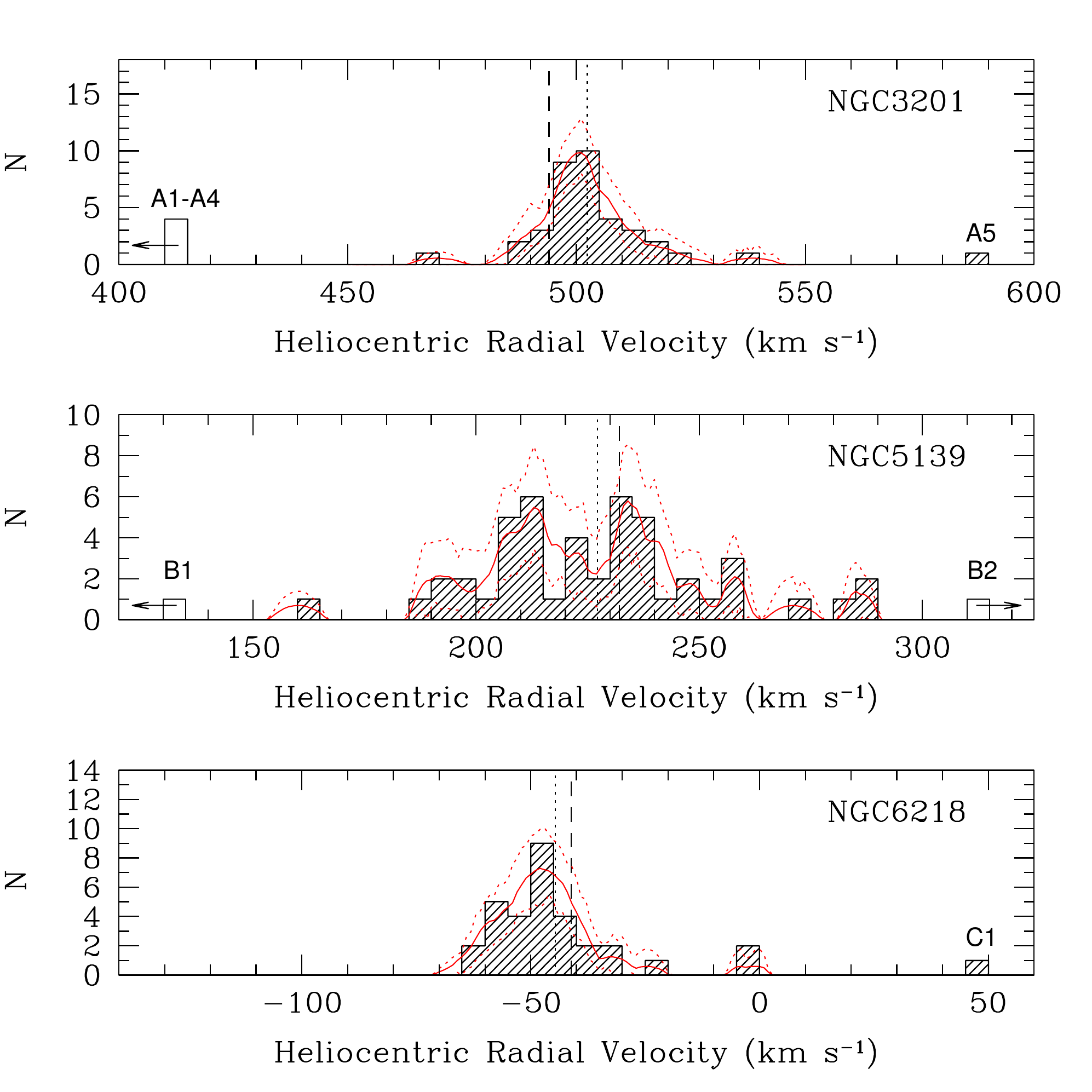}
\caption{Radial velocity distributions of the BSS candidates in each GC. The long-dashed lines mark the values of the GC radial velocities given in \cite{harris96},~while the short-dashed vertical lines are the median values obtained from our data.~The solid red line shows a non-parametric Epanech\-nikov-kernel probability density estimate with 90\% confidence limits represented by the dotted red lines.~The BSSs that have odd velocities are labeled in the figure. The non-filled bins correspond to BSS candidates with odd velocities out of the limits of the figure.~A1, A2, A3 and A4 have velocities between 25 and 56 km s$^{-1}$. B1 and B2 have velocities of 15.6 and 360.6 km s$^{-1}$ respectively.}
\label{rad_vel_3}
\end{figure}

We perform radial velocity measurements for all our spectra in order to study the membership likelihood and to compare our measurements with the GC radial velocity values found in the literature.~We use the {\sc IRAF} task {\sc fxcor} \citep{tonry79} to compute radial velocities via Fourier cross correlation between the science data and a set of template spectra.~As templates we use the high-resolution ($R\!=\!42000$) spectra from the {\sc ELODIE} library \citep{prugniel01}.~In most cases, strong absorption regions like the Balmer lines and the Mg$b$ triplet region were used in the cross correlation and for all our targets we avoided spectral regions affected by cosmic-ray hits and the CCD inter-chip gaps.~The radial velocities obtained are however relative to the frame of reference of the template spectra used in the task {\sc fxcor}, therefore we repeat the process with our observed standards stars and compare the radial velocities obtained against those in the literature.~This way we can calculate the radial shifts of the templates themselves and use this to calibrate the values obtained for each BSS.~The final laboratory restframe radial velocity was calculated by taking the error-weighed mean of five measurements from different templates. We used the {\sc IRAF} task {\sc rvcorrect} to obtain the final heliocentric radial velocities by using the date, location, and sky coordinates of the observations to correct for the Earth's relative velocity against the Sun.~The final heliocentric radial-velocity distributions of our sample BSS candidates are shown in Figure~\ref{rad_vel_3} for each GC. 

The corresponding heliocentric radial velocities of NGC\,3201, $\omega$Cen, and NGC\,6218 from \cite{harris96} are shown as long-dashed lines while the mean values of our measurements are shown as short-dashed vertical lines. For all GCs, we find mean radial velocity values that are not more than $\sim$4-8 km s$^{-1}$ different from the literature values.~We use our mean radial velocities to study the cluster membership of each BSS candidate. These values are $502.4\pm11.2, 227.2\pm26$, and $-44.6\pm14.2$ km s$^{-1}$ for NGC\,3201, NGC\,5139, and NGC\,6218, respectively\footnote{The corresponding median values are 500.2, 225.5 and -47.3 km s$^{-1}$, respectively.}, where the errors correspond to 1$\sigma$.~We adopt a 3-$\sigma$ cut and label outliers rather than eliminating them from the sample, in order to individually check their membership likelihood.~In Figure~\ref{rad_vel_3}, all outliers are labeled and the non-filled bins show stars with radial velocities off the limits of the figure. After this cut, we are left with 36, 47 and 31 BSSs in NGC 3201, $\omega$Cen and NGC 6218 respectively. Overall we find that $\sim\!93\%$ of our BSS candidates show radial velocities consistent with the system velocity of their parent GC. This suggests that the field star contamination is low (few \%), indicating that BSSs are majorly exclusive elements of dense stellar environments as found in GCs.~The outliers found in each sample can be divided arbitrarily into two groups: the ones with radial velocity differences larger than 100 km s$^{-1}$ with respect to the mean value, and the ones in which the difference is less than that.~We choose to further include outlier stars from the second group under the criteria that these could be BSSs that formed in strong dynamical events which then transferred large amounts of momentum into kinetic energy. These are star A5 and C1 from NGC 3201 and NGC 6218, respectively, which will be also studied in order to check their role in the final dynamical distribution functions for each GC. 

The bulk of the distributions in Figure~\ref{rad_vel_3} is, in all cases, fairly symmetrically distributed around the mean value, with very few outliers.~The width of the distributions is similar for NGC\,3201 ($\sigma_{\rm RV}\!=\!11.2$ km\,s$^{-1}$) and NGC\,6218 ($\sigma_{\rm RV}\!=\!14.2$ km\,s$^{-1}$), while $\omega$Cen shows about a factor of 2 larger value, i.e.~$\sigma_{\rm RV}\!=\!26$ km\,s$^{-1}$.~This is qualitatively consistent with the central velocity dispersions reported in \cite{harris96}, i.e.~$\sigma_0\!=\!5.0$ (NGC\,3201), $16.8$ (NGC\,5139), and $4.5$ km s$^{-1}$ (NGC\,6218).~However, the radial velocity dispersions are systematically larger than the central parts of the constituent stellar populations in each GC, which interestingly could be explained by dynamical interactions such as those that might form a BSS, or by the fact that BSSs are likely to be part of binary systems, and therefore their radial velocities could be enhanced by orbital motion. This intriguing result requires confirmation with larger BSS samples testing whether BSSs have different radial velocity dispersion profiles than the rest of the GC stellar population. 

\subsubsection{Field Stars Contamination}\label{fieldstarssection}
The fact that we are including stars such as A5 and C1 into our BSS sample calls for a better understanding of how susceptible our sample is to field stars contamination. In order to get an approximate idea we have checked the radial velocity distributions of field stars using the galactic model of \cite{robin2003} and found that the field stars in the line of sight of NGC\,3201 and NGC\,5139 have radial velocity values ranging $-50\!<\!v_r\!<\!150$ km s$^{-1}$ and $-100\!<\!v_r\!<\!100$ km s$^{-1}$, respectively. These velocities are clearly distinct from the systemic velocities found for these two globular clusters, and, therefore, the probability of our BSSs being false positives is very low. This is also the case for outliers such as star A5 which is also highly inconsistent with the radial velocity distribution of field stars along that line of sight. For field stars along the NGC\,6218 line of sight we find model predicted radial velocities in the range $-150\!<\!v_r\!<\!150$ km s$^{-1}$ (which includes the radial velocity found for star C1), even with a peak in their distribution function around $-40$ km s$^{-1}$, which is very similar to what is found for the mean radial velocity of BSSs in NGC\,6218. Hence, the BSS radial velocity distribution in this globular cluster requires particular attention.~For this we have searched the proper motion data catalog of \cite{zloczewski12} and found measurements for 16 out of our 32 BSSs studied in NGC\,6218.~All of them show proper motions consistent with being cluster members. We then check the radial velocity distribution of these 16 BSSs and find that they populate the entire radial velocity distribution found for the entire sample of 32 BSSs.~A Kolmogorov-Smirnov test shows that the sample of proper-motion confirmed BSSs and the entire sample are likely to have identical radial velocity distributions with a p-value $\!>\!0.8$. This, of course, does not completely rule out the possibility of including field stars in our sample, yet if there happened to be such contaminants, these results suggest that they would consist of a minor fraction. We do not have proper motion information for star C1 and so a final conclusion cannot be drawn concerning its membership probability. We note that this is the innermost star in the sample, which alone favours the cluster membership option. The inclusion of A5 and C1 in the confirmed BSS sample must, therefore, be handled with care.

\subsection{BSS Rotational Velocities}

\subsubsection{pPXF: The Code}\label{ppxfsection}
Rotational velocities are obtained for all our sample star spectra by using the penalized pixel fitting algorithm called pPXF, which is an IDL program to extract the kinematics of stellar populations from integrated-light absorption-line spectra of galaxies. This software implements the penalized pixel-fitting method developed by \cite{cappellari04} as a way to recover the central velocity, $v$, and the standard deviation, $\sigma$, as well as high-order moments such as skewness and kurtosis from the line-of-sight velocity distribution (LOSVD) through a Gauss-Hermite expansion of the absorption-line profile of the following form:
\begin{equation}
\mathcal{L}(\nu)=\frac{e^{-(1/2)y^2}}{\sigma\sqrt{2\pi}}\left[1+\sum_{m=3}^{M}h_mH_m(y)  \right]
\end{equation}
where $\nu$ is the frequency, $y=(\nu-v)/\sigma$, $H_m$ are Hermite polynomials and $\mathcal{L}(\nu)$ is the LOSVD. The program uses "initial guess" input values of $v$ and $\sigma$ to convolve template spectra in order to fit the object spectra. The fitting consists of a nonlinear least-squares optimization of the parameters $v,\sigma,h_3,...,h_M$.~The code then uses a penalty function, derived from the deviation of the object line profile from a Gaussian, which is added to the $\chi^2$ of the fit.~Therefore, a penalized $\chi^2$ is calculated as:
\begin{equation}
\chi^2_p=\chi^2(1+\lambda^2\mathcal{D}^2)
\end{equation} 
where $\mathcal{D}^2$ is the penalty function, that according to the authors can be approximated by 
\begin{equation}
\mathcal{D}^2\approx\sum^M_{m=3}h_m^2,
\end{equation}
and $\lambda$, which is a \textit{bias} parameter chosen by the user.~The fitting is iterated until the variance is minimized and the fit converges.~This pPXF procedure predisposes the solution with a single Gaussian when the S/N is low, while it is able to recover the higher-order moments of the absorption line spectrum when the S/N is high. For more details on the algorithm the reader is referred to \cite{cappellari04}.

\subsubsection{pPXF: Understanding Systematics and Capabilities through Monte-Carlo Simulations}
\label{ln:systematics}

\begin{figure}[t!]
\centering
\includegraphics[width=8.9cm]{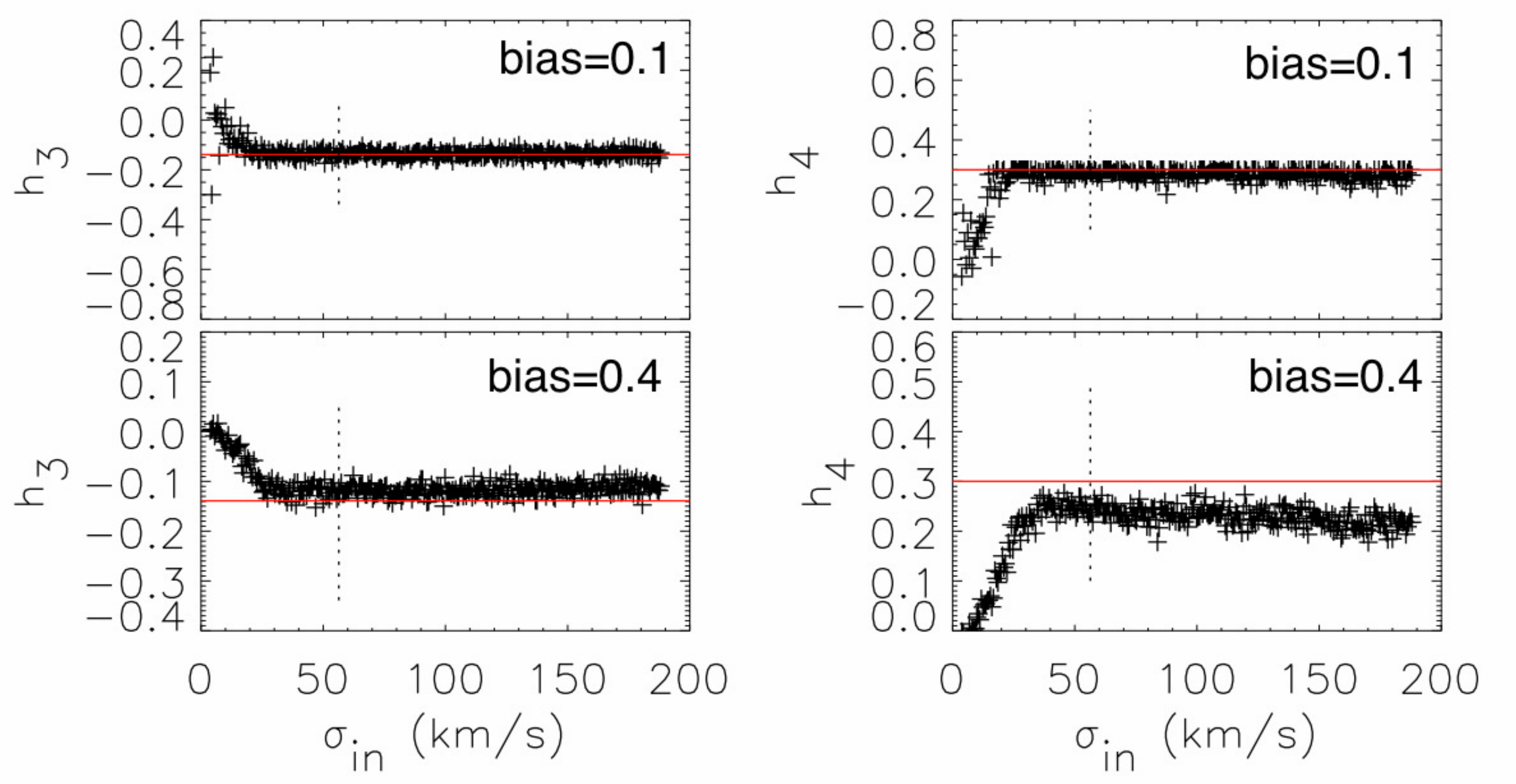}
\includegraphics[width=8.9cm]{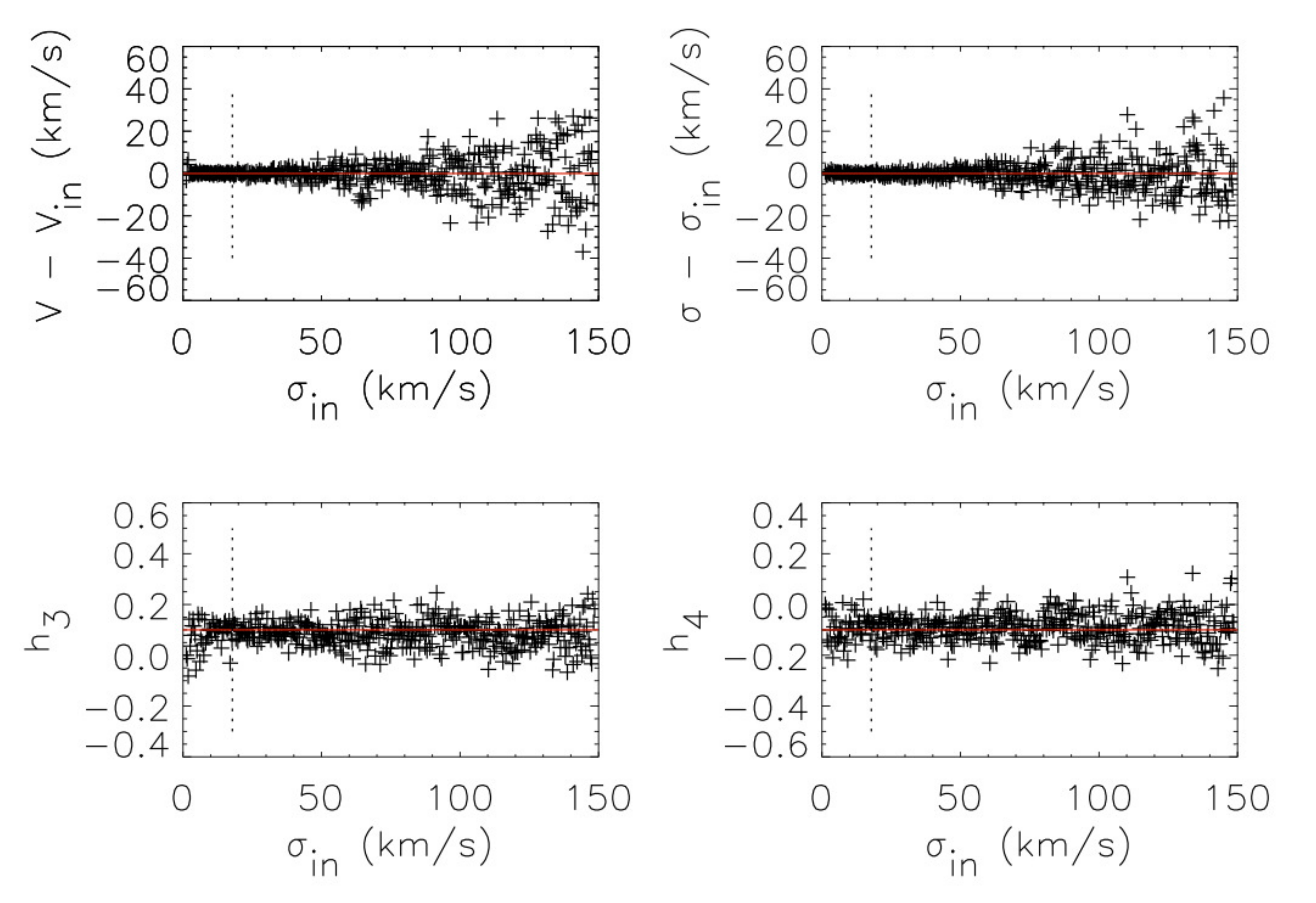}
\caption{({\it Top panels}): Monte-Carlo simulations using pPXF for estimating a proper \textit{bias} parameter. Plotted are the recovered Gauss-Hermite moments $h_3$ and $h_4$ as a function of the input velocity dispersion $\sigma_{\rm in}$. The upper panels show the results for a low value of the \textit{bias} parameter (0.1), while the bottom panels show the corresponding plots for a high \textit{bias} value (0.4). The code is very efficient in recovering the input values in the case of a low \textit{bias} parameter. The lower plots show the same experiment when using a high value for the \textit{bias} parameter and illustrate clearly that the algorithm underestimates the values of the high-order features if the penalty is too large.~In all cases the red line indicates the real value of the LOSVD and the vertical dashed lines mark the 3$\times$(velocity scale) limit, which is the point when the code starts becoming insensitive to any Gaussian deviation. ({\it Center and bottom panels}): Results of the measurements of the LOSVD parameters ($v, \sigma, h_3, h_4$) for different input values of $\sigma$. For all plots the red line marks the input values of the LOSVD and the vertical dashed line is the 3$\times$(velocity scale) limit, which is where the code starts becoming insensitive to any Gaussian deviation.}
\label{bias_test}
\end{figure}

The \textit{bias} parameter described above must be adjusted delicately to prevent the pPXF code from losing the high-order features in the object spectrum.~For this we preformed a series of Monte-Carlo simulations in which we used a representative spectrum from our data to inspect the behaviour of the penalty function. First we used pPXF to fit a LOSVD to the object spectrum using no penalty function (\textit{bias}=0) finding the un-biased values of $(h_3,h_4)$.~We then used these values as input and ran pPXF several times, each time convolving the object spectrum with a successively larger $\sigma_{\rm LOSVD}$ kernel.~We repeated this process for different values of the \textit{bias} parameter and found the maximum penalty value such that, for $\sigma > 3\times$(velocity scale), which is the lower limit at which the program becomes insensitive to any deviation from a Gaussian absorption profile at any S/N \citep{cappellari04}, the mean difference between the output and the input parameters is well within the scatter of the simulation. We then choose this maximum value as the \textit{bias} parameter. This is illustrated in Figure~\ref{bias_test} for two cases: one with a low penalty ({\it bias} = 0.1) and one with an overly high penalty ({\it bias} = 0.4). We observe that, as expected, a too large penalty can make the code lose information on the high-order features of the spectrum, such as $h_3$ and $h_4$. Our analysis shows that the value {\it bias} = 0.2 gives the best results. We use this {\it bias} value throughout all subsequent pPXF runs.

The next step is to test the stability of the solution when varying the amount of spectral range used by the fit. For this we create an artificial single Gaussian absorption line with a FWHM = 0.3\,\AA, which corresponds to the instrumental profile of the IMACS spectra, and add poisson noise to obtain S/N = 70 (see Figure~\ref{SNR_dist}).~We then perform the same simulation as above, i.e.~we convolve the absorption line to reproduce an arbitrary LOSVD and then try to recover the parameters using the same original artificial line as a template.~The key variable here is the amount of pixels (i.e.~continuum) that we include in the fit.~The spectral resolution of our resampled IMACS data is 0.1 \AA/pix.~Thus, a typical fitting range of 100 pixels corresponds to 10\,\AA.~We find that for different spectral ranges the code is able to recover $v$ and $\sigma$ to within about $\pm20$ km s$^{-1}$ when the absorption line is broader than $\sigma\!\ge\!120$ km s$^{-1}$, and to within about $\pm5$ km s$^{-1}$ when the absorption line has $\sigma\!<\!100$ km s$^{-1}$.~The bottom panels in Figure~\ref{bias_test} show the results from one of these simulations for a spectral range of 100\,\AA.~This result is stable for spectral ranges between approximately 70 and 300\,\AA.~For smaller and larger spectral ranges the code begins to produce uncertainties in $v$ and $\sigma$ close to $\pm40$ km s$^{-1}$ for absorption lines with $\sigma\!\geq\!100$ km s$^{-1}$.

\subsubsection{Measuring $v\sin(i)$ of BSS candidates}

\begin{figure}[t!]
\centering
\includegraphics[width=7.4cm]{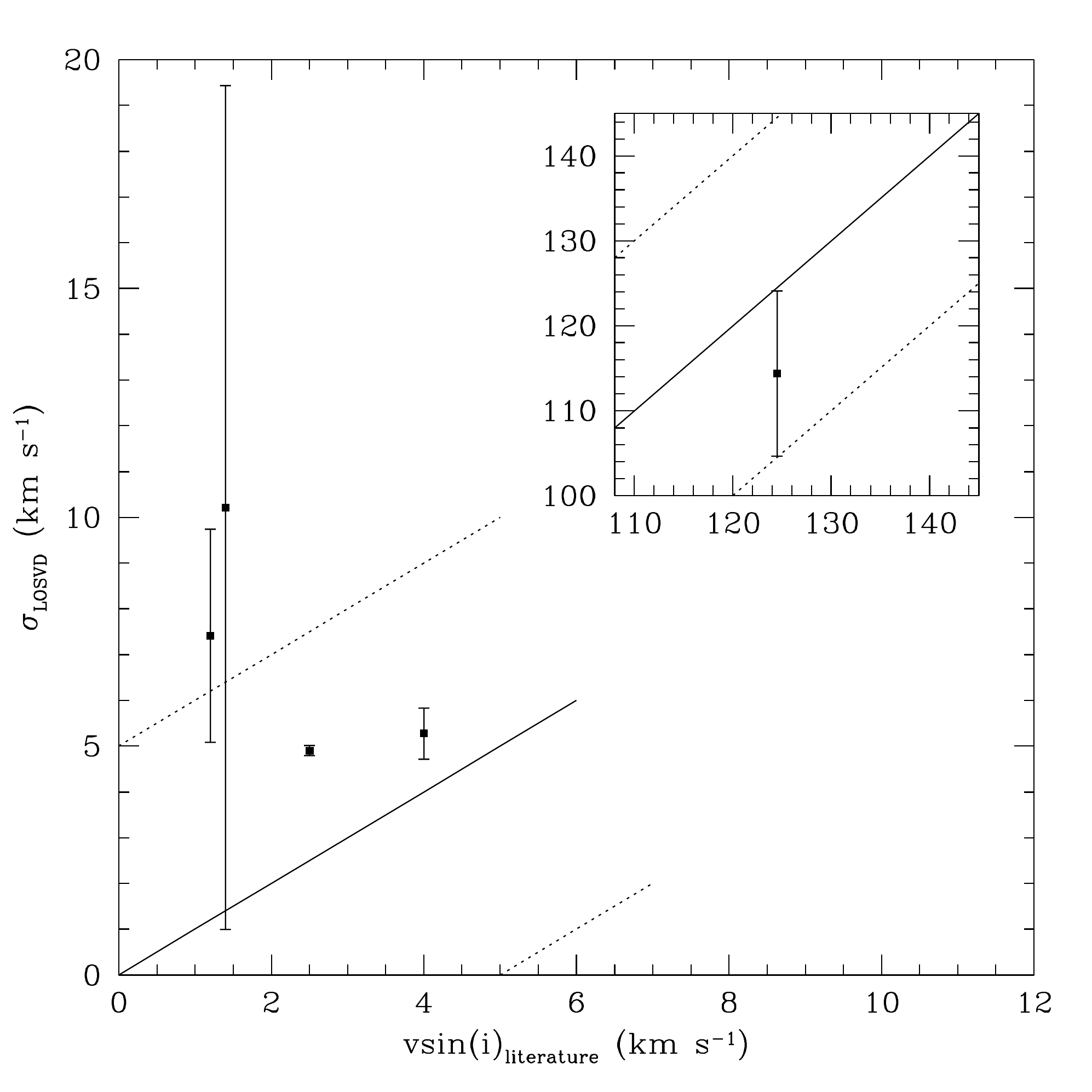}
\caption{Comparison of the $\sigma_{\rm LOSVD}$ values obtained with pPXF against literature values of $v\sin(i)$ for our rotational velocity standard stars. The solid line shows the unity relation and the dashed lines show the $\pm5$ km s$^{-1}$ typical error range found for low rotational velocities.~The small inset panel shows the same comparison for the fastest rotating standard star and the dashed lines show the $\pm\!20$ km s$^{-1}$ typical error range found for high rotational velocities, i.e.~$\sigma\!\ga\!100$ km s$^{-1}$.}
\label{stands_sigma}
\end{figure}

\begin{figure}[h!]
\centering
\includegraphics[width=7.39cm]{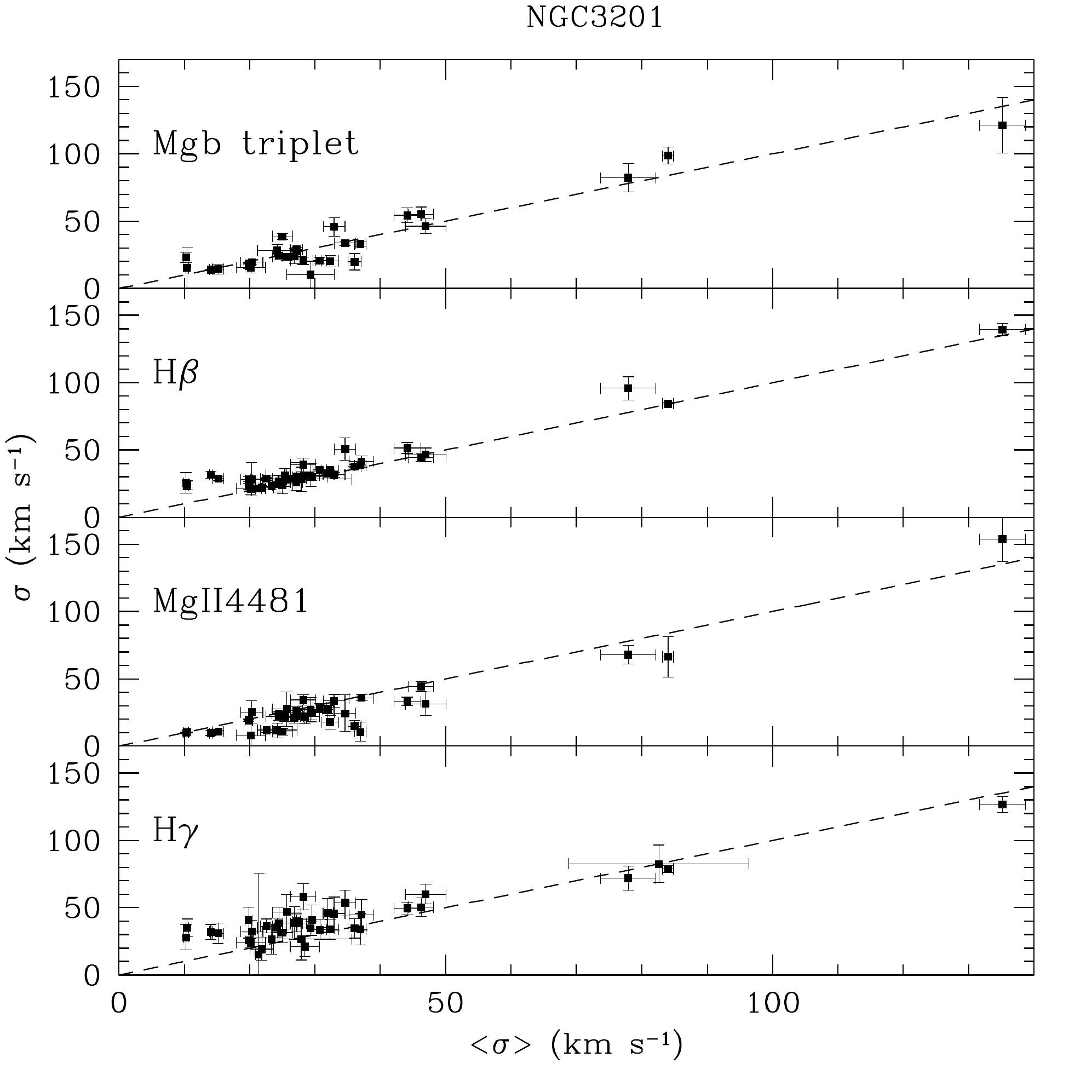}
\includegraphics[width=7.39cm]{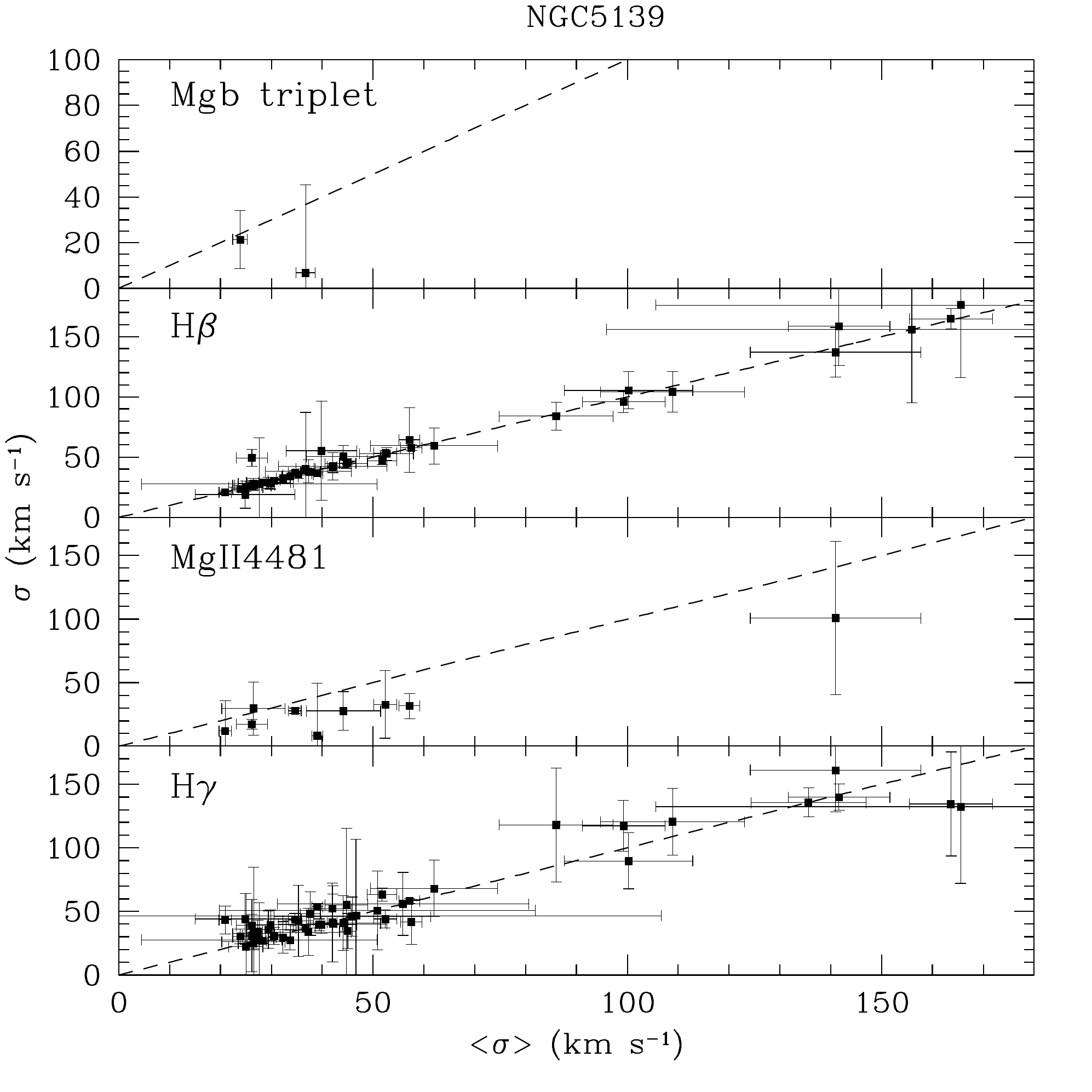}
\includegraphics[width=7.39cm]{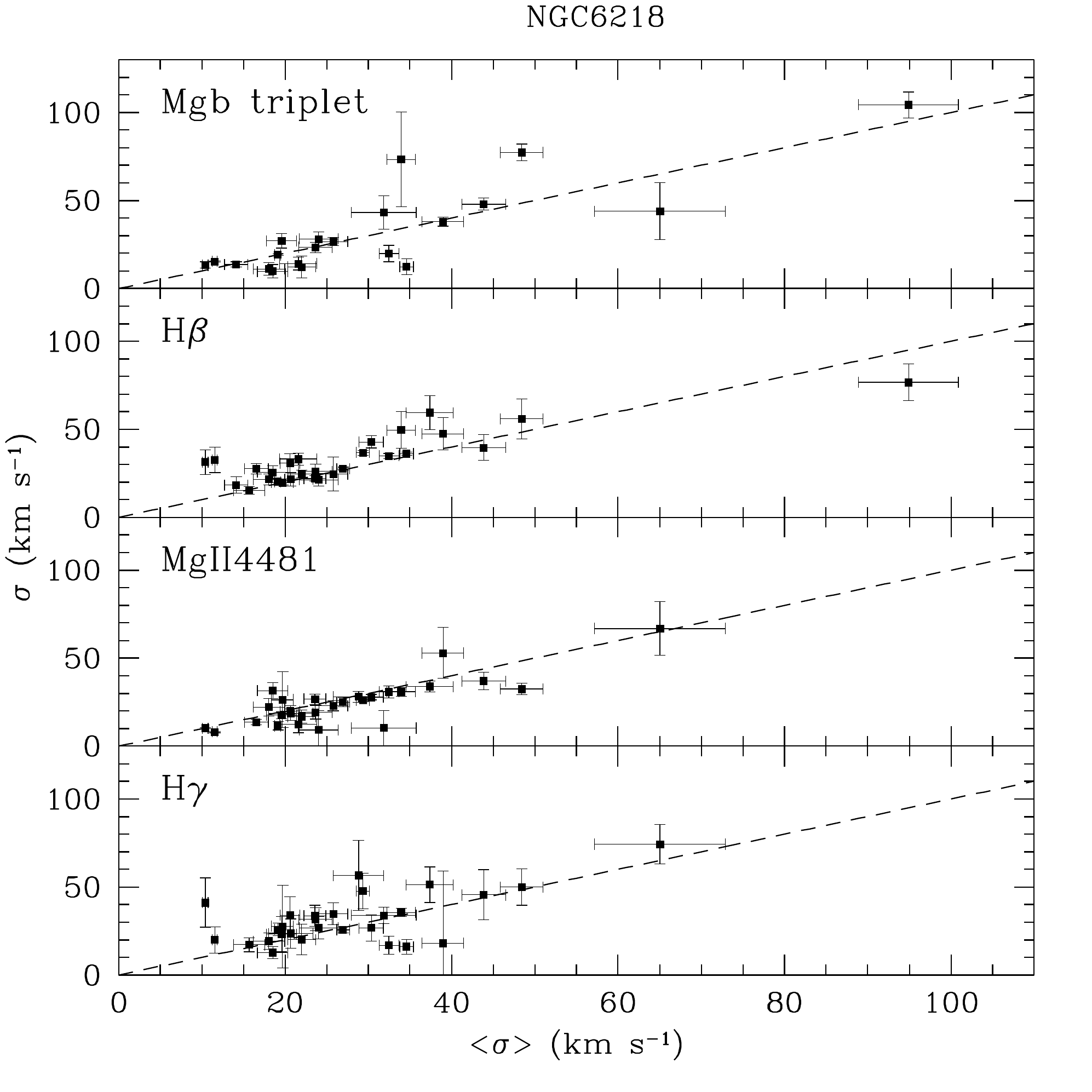}
\caption{Comparison of the final mean $\langle\sigma\rangle$ values versus the individual $\sigma$ measurements obtained at different spectral regions for all the BSSs candidates in NGC\,3201 ({\it top panel}), NGC\,5139 ({\it middle panel}), and NGC\,6218 ({\it bottom panel}).~The dashed lines show the unity relation in all panels.}
\label{sigmas_compare}
\end{figure}

The broadening of the absorption lines depends on the stellar rotational velocity and the viewing angle of the rotational axis.~Here we calculate the projected rotational velocities by adopting the relation:
\begin{equation}
v\sin(i) = \alpha\!\cdot\!\sigma_{\rm LOSVD}
\end{equation}
where $\sigma_{\rm LOSVD}$ is the standard deviation of the line-of-sight velocity distribution function obtained from pPXF and $\alpha$ is a proportionality factor. Although this factor cannot be measured from first principles and determined only statistically, we can calibrate the fidelity to recover the rotational velocities of standard stars.~Generally, differential effects related to variations in the micro/macro turbulence as well as temperature broadening are all assumed to be part of this scaling parameter $\alpha$, to ${\cal O}(0)$ independent of the input $\sigma_{\rm LOSVD}$.

\begin{deluxetable}{lrrr}[!ht]
\tabletypesize{\scriptsize}
\tablecaption{Sample properties of rotational velocity standards \label{tab:rotref}}
\tablewidth{0pt}
\tablehead{
\colhead{Star ID} &  \colhead{$v\sin(i)$} & \colhead{Spectral Type} & \colhead{Reference}
}
\startdata
HD69267   & 4.0   km/s         & K4III & \cite{fekel97} \\ 
HD71155   & 124.5 km/s         & A0V  & \cite{royer02} \\  &  &  &\cite{bernacca70} \\ 
HD131977 & 1.2   km/s         & K4V  & \cite{fekel97} \\ 
HD161096 & 2.5   km/s         & K2III & \cite{fekel97} \\ 
HD188512 & 1.4   km/s         & G8IV & \cite{fekel97}
\enddata
\tablecomments{These rotational velocity template standard stars were observed with the same instrumental setup and under the same observing conditions as our science targets.}
\end{deluxetable}

During our observations we obtained spectra of five rotational velocity standard stars which are summarized in Table~\ref{tab:rotref}.~The results of the calibration are illustrated in Figure~\ref{stands_sigma}.~Each measurement and error bar corresponds to the mean value and the 1-$\sigma$ value from four pPXF runs around different spectral regions including H$\beta$, H$\gamma$, the Mg$b$ triplet, and the MgII4481 line. Each of these four values are in turn the median value of ten pPXF runs which differ in the size of the spectral range around those absorption line features.~The dashed lines in Figure~\ref{stands_sigma} show the error range expected at these low and high values for $\sigma_{\rm LOSVD}$.~In particular we expect a dispersion of about 5 km s$^{-1}$ for low $\sigma_{\rm LOSVD}$ and a dispersion up to 20 km s$^{-1}$ for $\sigma_{\rm LOSVD}\!>\!100$ km s$^{-1}$ (see Section~\ref{ln:systematics}). We see that, within the errors, our measurements are consistent with the unity relation and therefore, for simplicity, we will adopt $\alpha\!=\!1$ and consequently, for the rest of this paper, we will consider our final $\sigma_{\rm LOSVD}$ measurements equal to $v\sin(i)$.

\begin{figure}[t!]
\centering
\includegraphics[width=8.9cm]{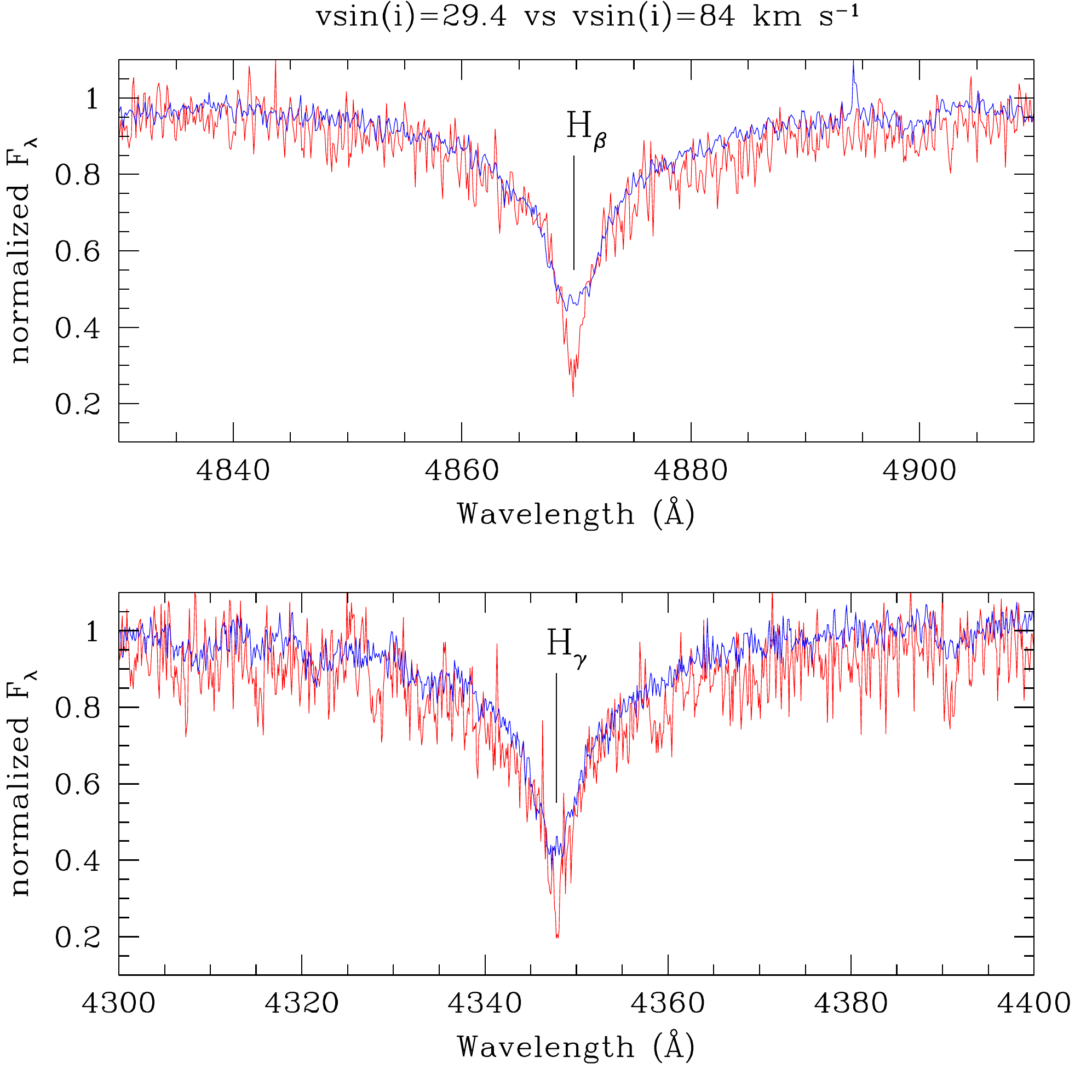}
\caption{Comparison between the spectral profiles of H$\beta$ and H$\gamma$ for two BSSs in NGC\,3201 with similar atmospheric parameters, particularly BSS2, in blue, and BSS19, in red, as labeled in Table~\ref{tab:n3201}, where the former has an estimated $v\sin(i)$=84.0 km s$^{-1}$ and the latter shows $v\sin(i)$=29.4 km s$^{-1}$. The difference in photometric color is $\Delta(V\!-\!I)$=0.04 mag, therefore implying very similar effective temperatures.~Note how the finer features are smoothed with fast rotation, as well as the bottom of the Balmer lines.}
\label{comp_balmer}
\end{figure}

For our BSS sample we run pPXF using spectra from the ELODIE high-resolution spectral library as templates. We select from this library twelve spectra with high S/N ($>\!100$) and spectral types A0V, F0V and G2V in order to sample an appropriate spectral type, i.e.~stellar mass range. The selection also requires the templates to have small rotational velocities ($v\sin(i)\!\simeq\!5$ km s$^{-1}$) in order to avoid an additive bias in the $v\sin(i)$ values and to guarantee enough sensitivity in the slow rotator range. In the same way, we also include template stars with moderate ($v\sin(i)\!\la\!30$ km s$^{-1}$), fast ($v\sin(i)\!>\!50$ km s$^{-1}$), and extremely fast ($v\sin(i)\!>\!100$ km s$^{-1}$) rotational velocities in order to properly fit BSSs with high rotational velocities. Similar to what is described for the standards, we calculate the final $v\sin(i)$ for a star by taking the error-weighed average $\sigma$ value of the fit of four, or less, spectral regions, depending on each spectra due to chip gaps and the quality of the fit. Each of these values is in turn a median value from ten fits, each of these having different sizes of the spectral range typically between 100 and 300 \AA. We plot the measurements from the fittings around each spectral region and compare them with the mean final value, $\langle\sigma\rangle$, to check how well constrained our results are and also to test for any strong dependence with the spectral region of the fit. This comparison is shown in Figure~\ref{sigmas_compare} for each BSS in each GC. The error bars on the y-axis come from taking the 1-$\sigma$ deviation in the set of ten values. Given that the wavelength ranges of the fit change in each run, therefore varying the amount of potentially bad pixels (specially close to the chip gaps), we are vulnerable to spurious solutions, which most of the times are left out when taking the median, but that affects the overall statistical uncertainty.~The large error bars in $\omega$Cen are mainly due to this effect, since these spectra are of the lowest S/N given the high level of crowding and difficult sky subtraction in each slit. For the other two GCs, however, we note that the expected errors can account for most of the dispersion between different measurements, and also the unity relation seems to hold for all spectral ranges in all GCs. This, in turn, rules out the possibility of any strong bias in $v\sin(i)$ introduced by the fit of a particular spectral region.~We interpret this last result also in favor of including regions around the Balmer lines as acceptable $v\sin(i)$ estimators even though the Balmer lines themselves are very sensitive to gravity and temperature variations.~In fact, we find that if we select a pair of BSSs from the same GC which have very similar colors (therefore we assume similar metallicity and temperature) and different estimated rotational velocities, the spectral profile of the H$\beta$ and H$\gamma$ lines are almost identical between a fast and a slow rotating BSS, except for the shape of the bottom of the line, which gets smoothed with fast rotation (see Figure~\ref{comp_balmer}).~The fitting will be, therefore, mostly determined by the finer absorption features and by the shape of the bottom of the Balmer lines.~As a final check, we note that the MgII4481 line should be ideal for $v\sin(i)$ measurements in late A type stars given that it is free from strong pressure broadening \citep{gray08}.~Therefore, the consistency observed in Figure~\ref{sigmas_compare} is suggesting that, within the errors, we can use the regions around the Balmer lines as reliable indicators.

\begin{figure}[t!]
\centering
\includegraphics[width=8.9cm]{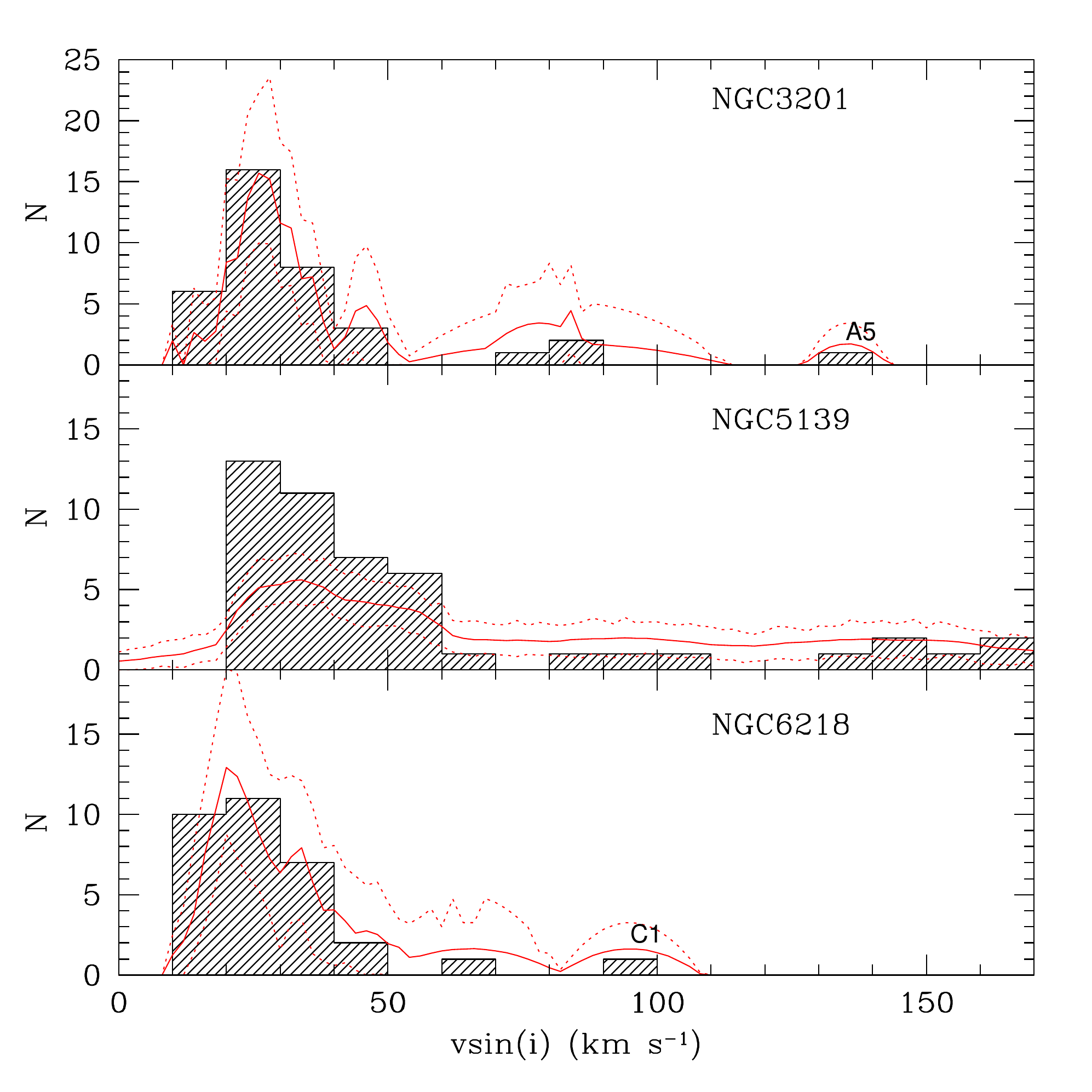}
\caption{Distributions of the $v\sin(i)$ measurements for all BSSs in NGC\,3201 ({\it top}), NGC\,5139 ({\it middle}), and NGC\,6218 ({\it bottom panel}).~The red solid curves illustrate non-parametric probability density estimates using an Epanechnikov kernel function together with their 90\% confidence limits shown as dotted curves. Note that the probability density estimate takes into account the uncertainties of the individual measurements. Due to the lower S/N of the NGC\,5139 measurements the curves appear less peaked than for the other two GCs. The two labeled BSSs (A5 and C1) with extreme $v\sin(i)$ values correspond to those labeled in Figure~\ref{rad_vel_3}.}
\label{rot_dist_3}
\end{figure}

The final $v\sin(i)$ distribution for each GC is shown in Figure~\ref{rot_dist_3}, while the values are listed in Table~\ref{tab:n3201}, Table~\ref{tab:n5139} and Table~\ref{tab:n6218}, for NGC\,3201, $\omega$Cen and NGC\,6218, respectively. The cases of NGC\,3201 and NGC\,6218 are similar, in the way that both show the peak of the distribution around $\sim\!20\!-\!30$ km s$^{-1}$. Approximately 90\% of the distributions for these GCs are found in the range of between 10 km s$^{-1}$ and 50 km s$^{-1}$. The other 10\% of BSSs, which have $v\sin(i)\!>\!50$ km s$^{-1}$, lie in the long tail of the distribution which goes up to $\sim\!90$ km s$^{-1}$ for NGC\,6218 and up to $\sim\!130$ km s$^{-1}$ for NGC\,3201. Note that in these GCs, the fastest rotating BSSs are outliers in the radial velocity distribution, which could also be a clue to their formation history. The case of $\omega$Cen is different since it shows a $v\sin(i)$ distribution slightly shifted towards higher rotational velocities. The bulk of its distribution function is between 20 and 70 km s$^{-1}$, with a peak around 30 km s$^{-1}$. This bulk contains 80\% of the total sample and the remaining BSSs form a tail of the distribution function with $v\sin(i)$ values up to $\sim\!$160 km s$^{-1}$. For this massive GC, the fraction of BSSs with $v\sin(i)\!>\!50$ km s$^{-1}$ corresponds to 30\%, in agreement to what was reported in \cite{lovisi13a} for their sample.

We note that, contrary to what is normally found, we find no BSSs rotating slower than 10 km s$^{-1}$, and, in the case of $\omega$Cen, no BSSs rotating slower than 20 km s$^{-1}$. Therefore, we may be systematically overestimating the $v\sin(i)$ for slow rotators.~One possible explanation of such a relatively high lower $v\sin(i)$ cut-off might be simply due to our sample selection. Another reason might be that our $v\sin(i)$ calibration (Figure~\ref{stands_sigma}) which is based on late-type stars to fix the slow rotating end of the $v\sin(i)$ parameter space, is affected significantly different than in the hotter target BSSs, probably due to large differences in the micro and macro-turbulence, as well as in the temperature broadening between late and early type stars, which could well be of the order of 10 km s$^{-1}$.~Since we concentrate on distribution functions of slow and fast rotators and only care about distinguishing between them, we do not expect this issue to affect any of our conclusions based on the differential analysis.

\section{Discussion}

\subsection{BSS Spatial Distribution}
One of the strongest clues to understanding the nature of BSSs comes from their spatial distribution in GCs. The radial distribution profiles of the BSS fractions are clearly bimodal for many GCs \citep{ferraro97}, but much flatter for others \citep{dalessandro08,ferraro06b}.~This radial density profile morphology was recently claimed to be tightly related to the dynamical state of the parent GC \citep{ferraro12}.~At the same time, many ideas were put forward regarding the dependence of this distribution on the environment for different BSS formation channels \citep{piotto04, davies04}. In particular, it was claimed that mass transfer BSSs would form preferentially in the loose outskirts of GCs where the low stellar densities would allow binary systems to survive long enough to transfer sufficient stellar material from the donor star, while such binaries would get disrupted much more rapidly in the dense cluster cores. In contrast, BSSs formed through collisions were expected to be found preferentially in GC cores, where the stellar densities are high enough for these relatively rare events to actually occur at significant rates \citep{ferraro97}.~This scenario has not yet clearly been demonstrated observationally to be at work, and evidence so far is very unclear about the spatially dependent importance of the different BSS formation channels due to the fact that both types of BSS populations are simultaneously present in a GC.

\begin{figure}[!t]
\centering
\includegraphics[width=8.5cm]{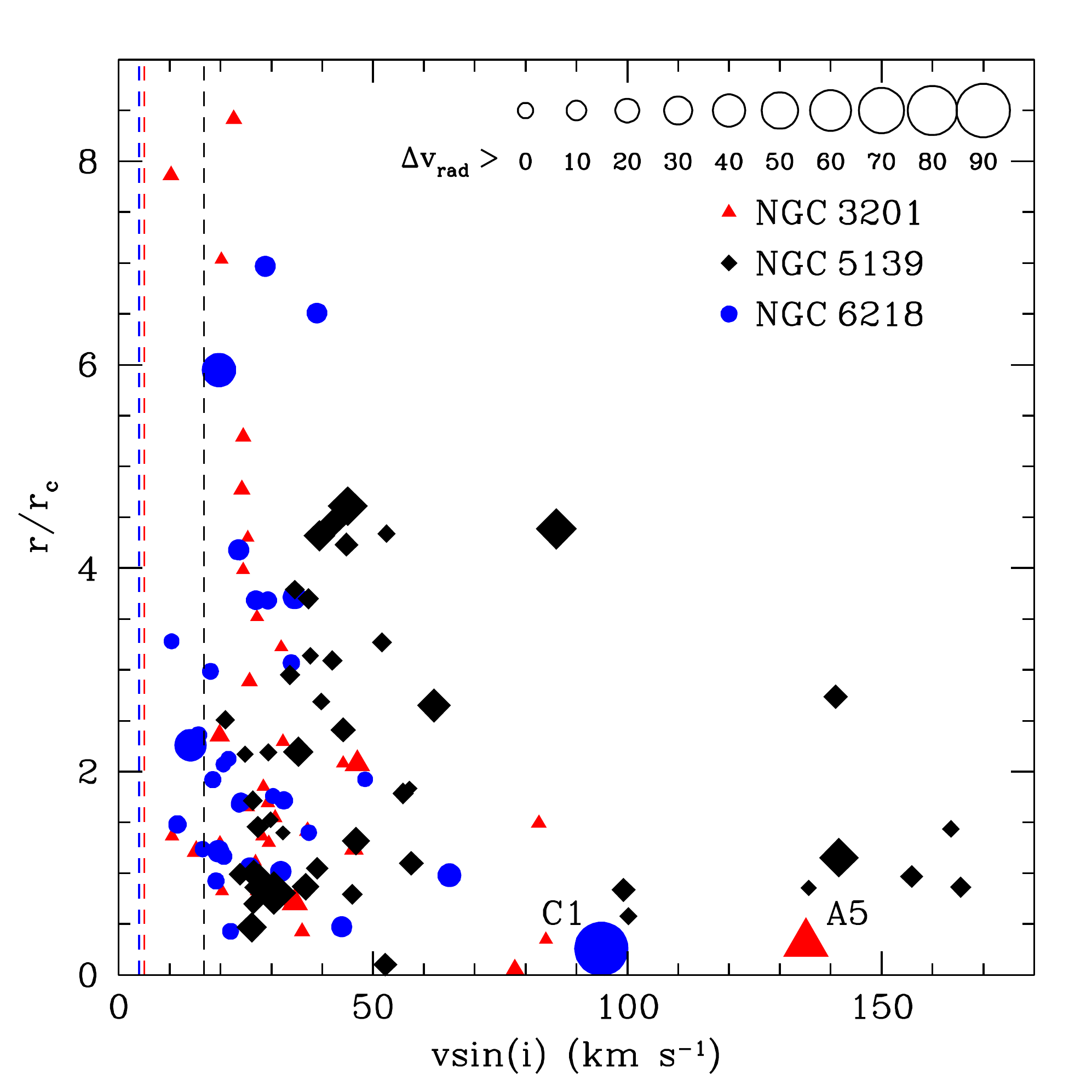}
\caption{Correlation between the projected cluster-centric distance vs.~$v\sin(i)$ for BSSs in NGC\,3201 ({\it red triangles}), NGC\,5139 ({\it black diamonds}), and NGC\,6218 ({\it blue squares}). The symbol sizes are parametrized by the radial velocity offset with respect to the systemic velocity of the host GC and are scaled as shown in the upper right. The dashed vertical lines indicate the central velocity dispersion of each GC, color coded as the symbol color. The two labeled BSSs correspond to the ones in Figures~\ref{rad_vel_3} and \ref{rot_dist_3}. Note that for NGC\,5139 we lack spatial coverage beyond $\sim\!4.5\,r_c$ and, thus, do not cover the outermost regions (see also Figure~\ref{cmd1}).}
\label{vsini_dist}
\end{figure}

The large spatial coverage of our BSS sample allows us to search for spatial correlations between $v\sin(i)$ and the cluster-centric radius.~This may unveil differences in the two populations identified by their $v\sin(i)$ signatures: BSSs that belong to the bulk of the $v\sin(i)$ distribution, i.e.~with rotational velocities around $\sim\!10\!-\!70$ km s$^{-1}$ and rapidly rotating\footnote{The term 'fast rotator'  will, from now on, be used for BSSs with $v\sin(i)\!\ga \!70$ km s$^{-1}$, unless it is stated otherwise.} BSSs with $v\sin(i)\!>\!70$ km s$^{-1}$.~In Figure~\ref{vsini_dist} we plot the BSS $v\sin(i)$ values against their projected cluster-centric distance in units of the GC core radius, adopting the values $r_c\!=\!1.3\arcmin, 2.37\arcmin$, and $0.79\arcmin$ for the core radius of NGC\,3201, NGC\,5139, and NGC\,6218, respectively, taken from the 2010 update of \cite{harris96}.~We find that, with the exception of two BSSs in $\omega$Cen, all fast rotating BSSs are significantly concentrated within $\sim\!2\,r_c$, while BSSs with $v\sin(i)\!<\!70$ km s$^{-1}$ populate the entire spatial extent of their parent GC. This relation also holds for BSSs with $v\sin(i)$ below and above 50 km s$^{-1}$ (the usual fast rotator limit) in NGC 3201 and NGC 6218.~We also show, as vertical dashed lines, the central velocity dispersion of each GC in order to asses the possibility of star blending causing the inner BSSs to appear artificially as fast rotators.~We see that, in all three GCs, the velocity dispersion is significantly smaller than the rotational velocities of the fastest BSSs and therefore unable to be causing this spatial pattern.~Interestingly, the fastest rotators in NGC\,3201 (A5) and NGC\,6218 (C1) have anomalous radial velocities and are located, in projection, even deeper in the cluster center at $<\!0.5\,r_c$. Like the two fast rotators in $\omega$Cen at $>\!2\,r_c$, these two rapidly rotating BSSs could possibly be in the early stage of ejection from their host GCs, given their projected locations and radial velocity offsets with respect to their parent stellar system. On the other hand, if the two fast rotators in $\omega$Cen at $>\!2\,r_c$ were also formed in the inner regions, then they are possibly already well in the process of ejection. In agreement with this, we find that the furthest out of the two has a radial velocity of 288.3 km s$^{-1}$, right on the edge of the radial velocity distribution of $\omega$Cen (see Figure~\ref{rad_vel_3}). Another interpretation is simply that fast rotating BSSs form at the same rate both in the outskirts and in the central core of its parent GC, and they later sink down into the inner regions due to mass segregation processes.~Our data favors the first scenario, the one in which fast rotating BSSs form preferentially in the inner regions, since, in all three GCs, we find slow rotating BSSs distributed across the entire cluster's spatial extent with a certain underlying density profile, while fast rotators seem to depart from such a profile into a much more concentrated one. A Kolmogorov-Smirnov test has been performed on the slow- and fast-rotating BSS subsamples and it was found that the hypothesis in which the group consisting of BSSs with $v\sin(i)\!>\!70$ km s$^{-1}$ and the one with $v\sin(i)\!<\!70$ km s$^{-1}$ are drawn from the same cluster-centric density distribution has a p-value $\!<\!0.005$. This is consistent with our idea that fast rotating BSSs are indeed located preferentially in the deeper regions of these clusters. The same test was performed for the individual clusters and we found a p-value $\!<\!0.04$ for NGC\,3201 and NGC\,6218, while NGC\,5139 gives a p-value $\!<\!0.2$. Therefore, even for the individual samples we find that the radial distributions of fast rotating BSSs are consistent with being shifted towards the inner regions compared to the radial distribution of slow rotating BSSs.

\begin{figure}[!t]
\centering
\includegraphics[width=8.9cm]{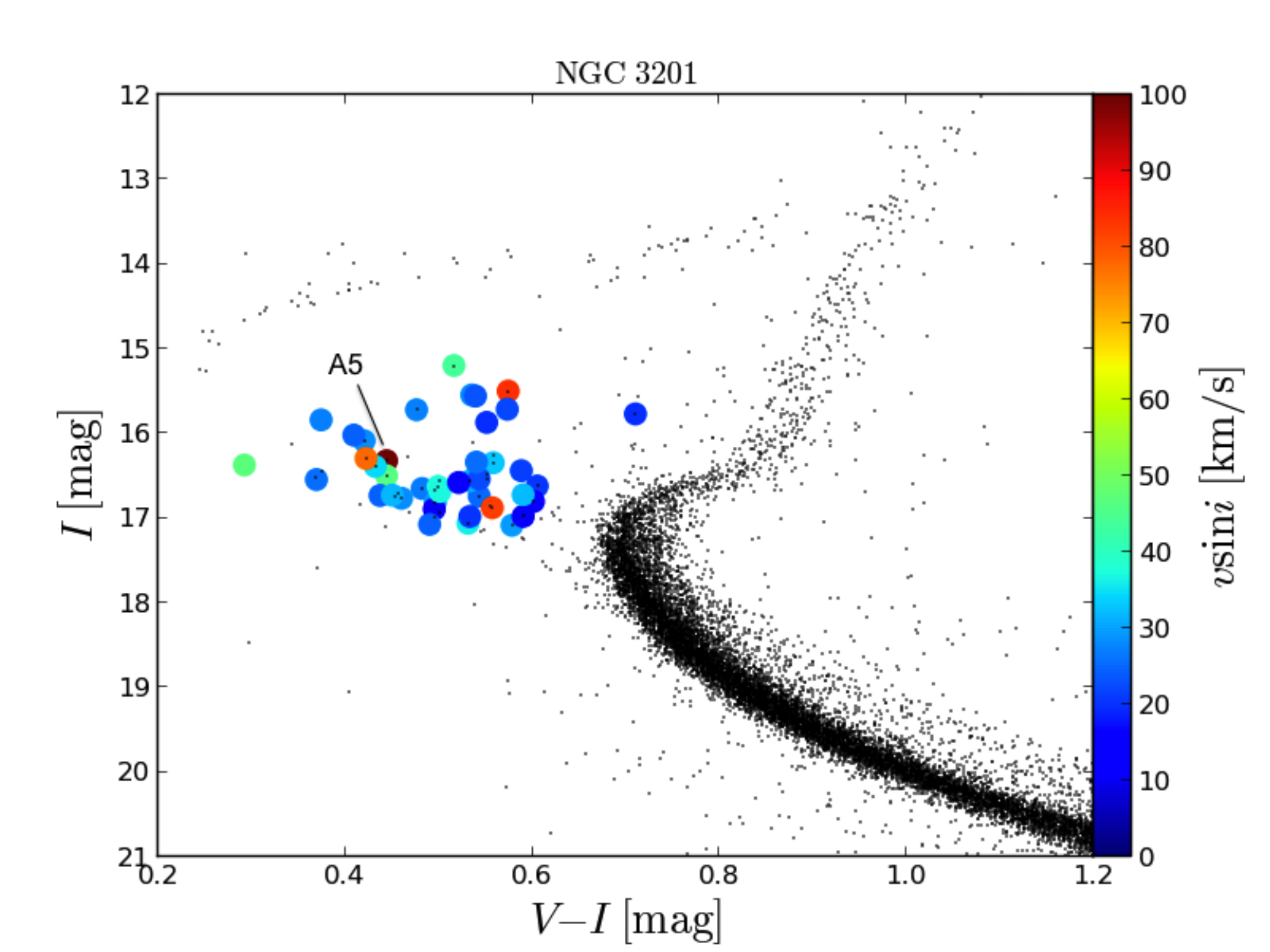}
\includegraphics[width=8.9cm]{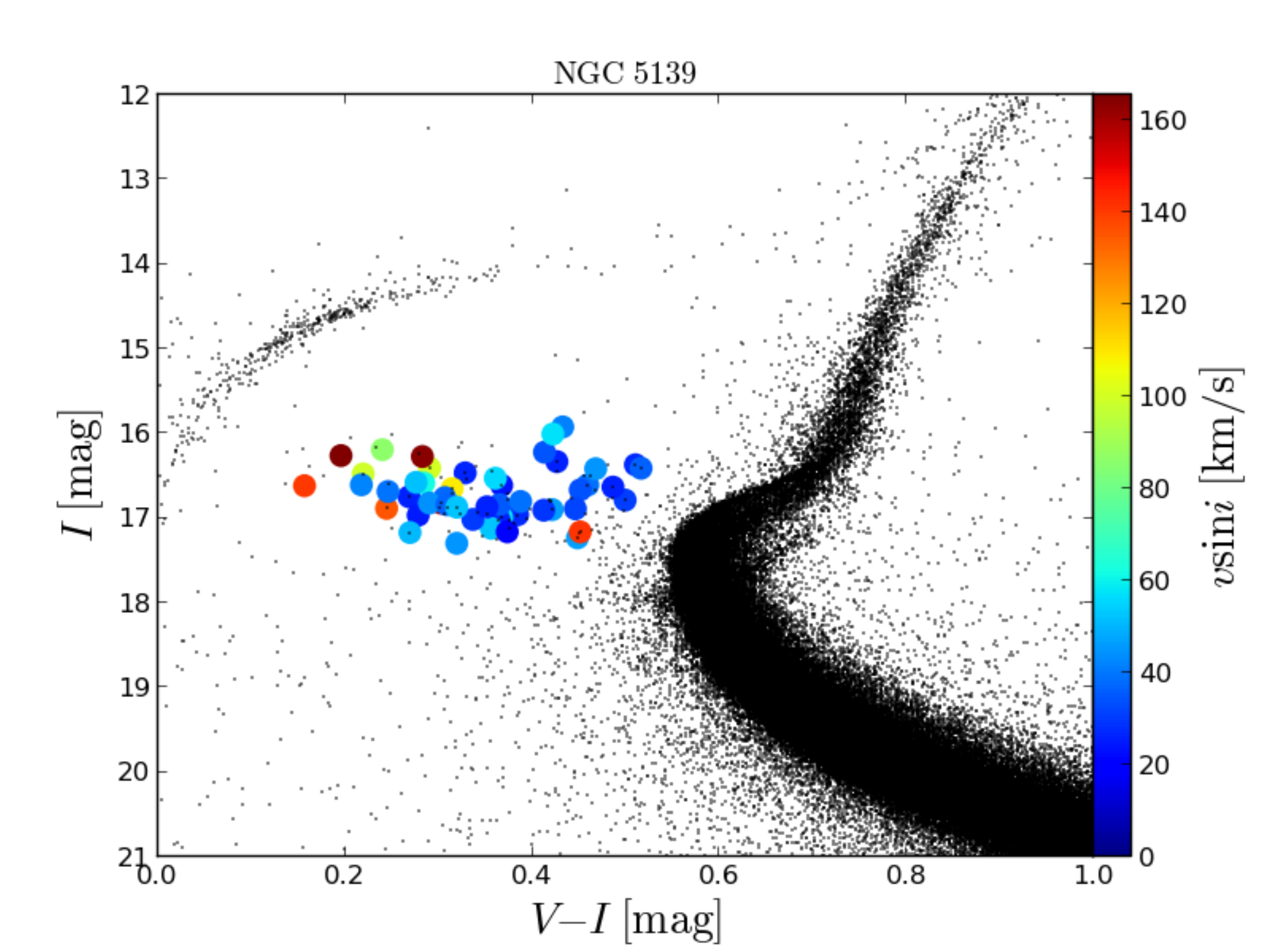}
\includegraphics[width=8.9cm]{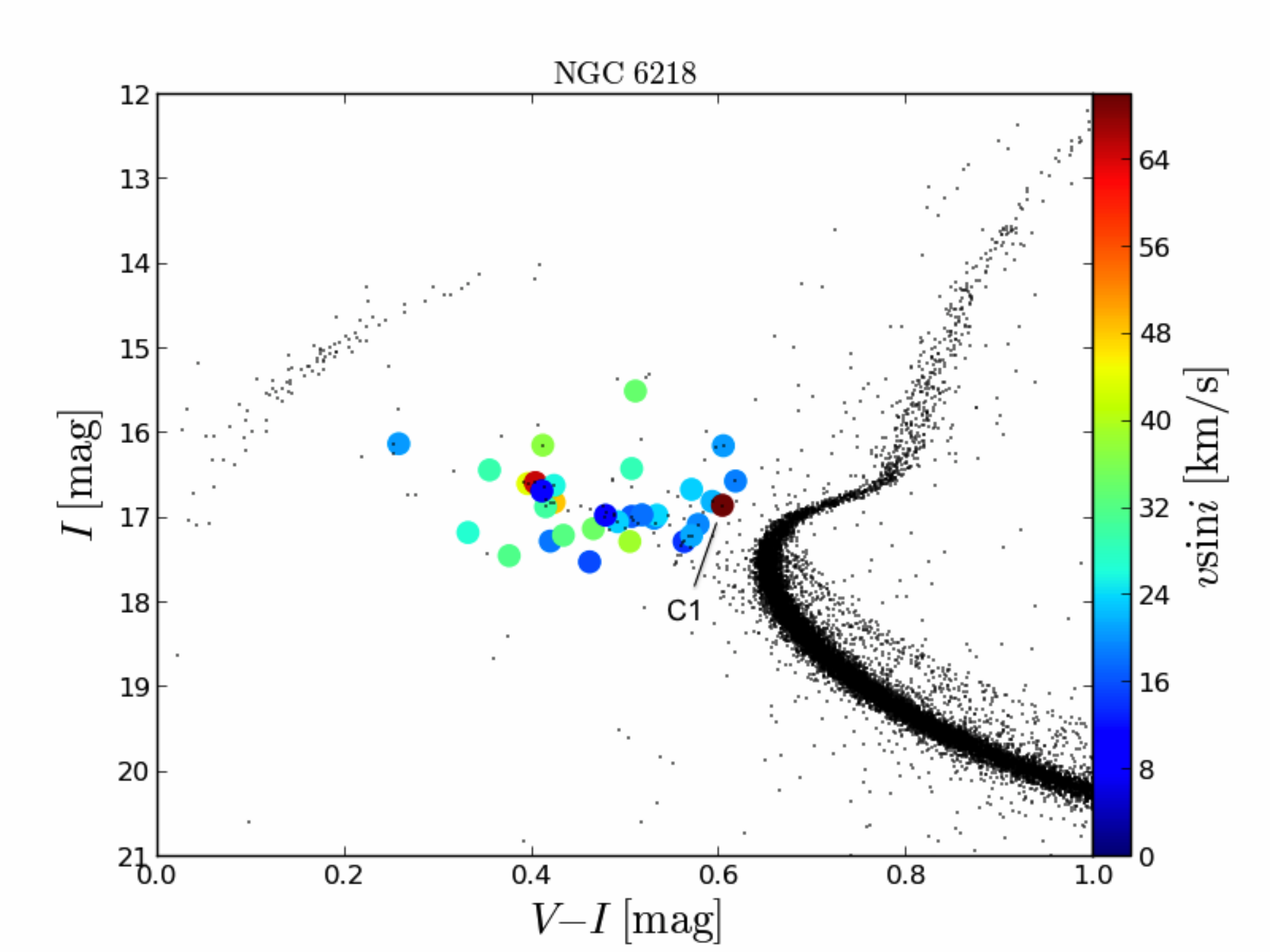}
\caption{Color-Magnitude Diagrams for all three GCs. Big circles are the BSS candidates and the color of the circle shows the $v\sin(i)$ value. Note that bluer BSSs are preferentially faster rotators than redder BSSs. This is best seen in $\omega$ Cen. The BSS with anomalously high radial velocities corresponding to those labeled in Figure~\ref{rad_vel_3} and \ref{rot_dist_3} are labeled accordingly.}
\label{cmd_vsini}
\end{figure}

\cite{lovisi10} reported three fast rotators in M4 having anomalous radial velocities and suggested a scenario in which three and four-body interactions occurring in the dense GC cores may be responsible for creating these fast rotating BSSs by transferring large amounts of angular momentum into kinetic energy.~As stated in the last paragraph, we think our results can be linked to the same phenomenon, where the progenitors of some fast-rotating BSSs in our sample experienced a recent interaction event in the GC core regions that put them on a hyperbolic trajectory, kicking the BSS out of the GC gravitational potential at relatively high velocities ($v_{\rm rad}\!\sim\!100$ km s$^{-1}$). 

\subsection{Spin-Down and Ejection Timescale Estimates for Rapidly Rotating BSSs}
As will be show in a future paper of this series, we find typical BSS masses in the range of $0.8\!-\!1.35\,M_\odot$ in our sample GCs.~We use the stellar evolution models of \cite{ekstroem12} to estimate the typical spin-down timescales for stars in this mass range and find that the longest spin-down time to reach equatorial rotational velocities below $v_{\rm eq, rot}\simeq10$ km s$^{-1}$ is about $200\!-\!300$ Myr for $0.8\,M_\odot$ stars and decreases to $\sim\!100$ Myr for a $1.35\,M_\odot$ star. In addition to this internal stellar dynamics estimate, we use a back of the envelope calculation including the quantities provided in \cite{georgiev09} and \cite{harris96} to compute the escape times from the GC centre out to the tidal radius for the two rapidly rotating BSSs in NGC\,3201 (A5) and NGC\,6218 (C1). Our calculations give an upper limit of $\sim\!400$ kyr for these BSSs to escape their parent GCs, which would mean that these two rapidly rotating BSSs, if indeed were formed in the core, must have been formed no longer than some hundred thousands years ago, assuming also that the dynamical interaction which set them at high velocities was the same that caused their formation.~Similarly, the BSS in $\omega$Cen located at $>\!2\,r_c$ and with high relative radial velocity previously mentioned could have migrated to its projected location after $\sim$200 kyr if it started its ejection at $\sim$1$r_c$.~Hence, we suggest that these rapidly rotating BSSs that are likely in the process of ejection must have experienced strong dynamical interaction no longer than $\sim$10$^5$ yrs, which would set a very significant constraint on their age, if we assume that these dynamical interaction were also responsible for their formation.~More generally, based on the spin-down timescale estimates, we suggest that the observed rapidly rotating BSSs with $v\sin(i)\!>\!70$ km s$^{-1}$ formed in strong dynamical interaction events in the central regions of their host GCs no longer than $\sim\!300$ Myr ago.

\subsection{BSS colors vs.~$v\sin(i)$}
\begin{figure}[t!]
\centering
\includegraphics[width=8.5cm]{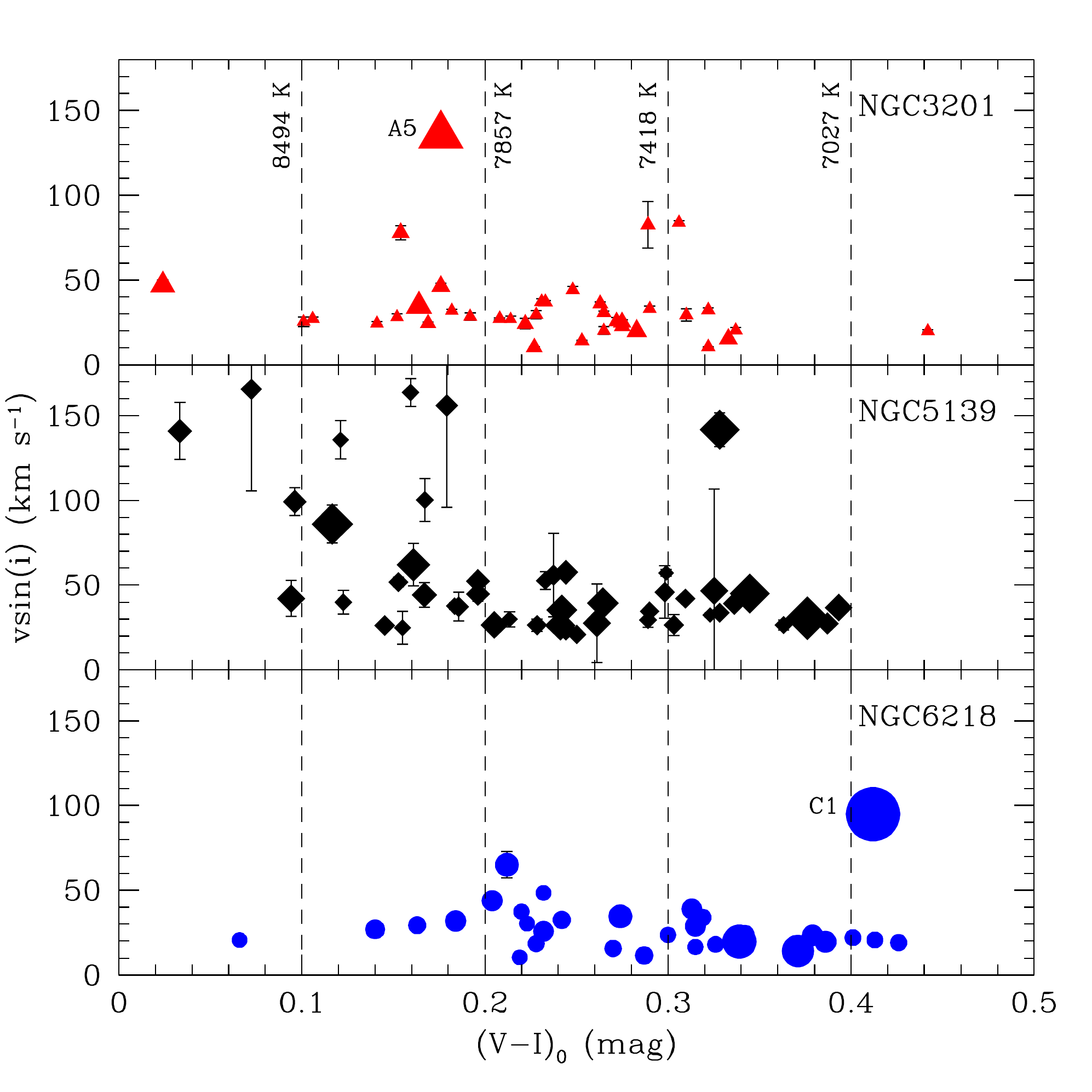}
\caption{Dereddened $V\!-\!I$ colors versus $v\sin(i)$ values for our sample BSSs in each target GC, NGC\,3201, NGC\,5139, and NGC\,6218. Note significant increase of the $v\sin(i)$ dispersion in NGC\,5139 for BSS colors $(V\!-\!I)_0\la0.25$ mag. The symbol sizes are parametrized by the radial velocity offset with respect to the systemic velocity of the host GC and are scaled as shown in Figure~\ref{vsini_dist}. The dashed vertical lines show stellar effective temperature values that correspond to different $V\!-\!I$ colors as derived by the relations from \cite{bessell98}.}
\label{color_vsini}
\end{figure}

The CMD location of BSSs in relation with their parent GC's stellar population can be a powerful tool for determining their formation history. In Figure~\ref{cmd_vsini} we plot the CMDs of our target GCs and explicitly show the BSS $v\sin(i)$ values in order to search for global trends. Even though slow and fast rotators are spread over all colors and magnitudes, an unexpected global trend seems to hold for NGC\,5139: the fastest rotating BSSs are preferentially bluer than the slower rotating BSSs.~We illustrate this more clearly in Figure~\ref{color_vsini} where we plot the BSS $v\sin(i)$ vs.~their dereddened $V\!-\!I$ color, i.e. $(V\!-\!I)_0$. Reddening values were obtained from \cite{dotter10} for NGC\,3201 and NGC\,6218 and from \cite{villanova07} for $\omega$Cen using \cite{cardelli89} extinction law with $R_V=3.1$.~The increase in the $v\sin(i)$ dispersion for BSS with $(V\!-\!I)_0\la0.25$ mag is striking.~In NGC\,3201 and NGC\,6218, the number of blue BSSs is smaller and a larger sample is required to decide whether a similar trend exists in these GCs. In any case, the majority of red BSSs with colours $(V\!-\!I)_0\ga0.25$ mag show a significantly smaller $v\sin(i)$ dispersion than their blue counterparts with few outliers at high $v\sin(i)$ values. We defer the discussion of correlations between $v\sin(i)$ values and stellar masses and ages to a future paper, but point out that there is an apparent $v\sin(i)$ dichotomy between cool and hot BSSs.

\section{Summary}
We obtained multi-object spectroscopy ($R\!\approx\!10000$) with IMACS on the 6.5-meter Baade Telescope at Las Campanas observatory for 137 BSS candidates in three Milky Way GCs (NGC\,3201, NGC\,5139, and NGC\,6218). The BSS candidates were selected from optical HST/ACS and ESO/WFI photometry and resulted in 116 ($\sim\!93\%$) of confirmed BSSs with radial velocities consistent with the host GC systemic velocity  for which good quality spectra could be obtained. We convolve template spectra to fit the absorption line profiles of several strong spectral features employing the pPXF technique \citep{cappellari04} and conduct detailed Monte-Carlo simulations to determine the fidelity and to constrain the influence of systematics involved in our analysis.

We find a bimodal distribution of BSS $v\sin(i)$ values in all three target GCs, with $\sim\!90\%$ of the BSS population having $v\sin(i)$ values between 10 and 50 km s$^{-1}$ and a peak value around $20\!-\!30$ km s$^{-1}$ in NGC\,3201 and NGC\,6218, while in $\omega$Cen $\sim\!80\%$ of the BSS population has $v\sin(i)$ values between 20 and 70 km s$^{-1}$ and a peak around 30 km s$^{-1}$.~The lower limit of the $v\sin(i)$ distributions (and hence, their peaks) seem higher than what is found in other GCs and we argue that this could be due to either our sample selection, or due to a systematical overestimation of $v\sin(i)$ for the slowest rotating BSSs by $\sim$10 km s$^{-1}$.~This, in any case, would not affect our conclusions based on comparative analysis (fast versus slow rotating BSSs).~For all GCs, we find rapidly-rotating BSSs with $v\sin(i)\!>\!70$ km s$^{-1}$, which are predominantly found in the central regions of their parent GCs and have sometimes differential radial velocities that are consistent with stars in the process of being ejected from their host stellar systems through hyperbolic orbits.~We discuss the spin-down timescales of these rapidly-rotating BSSs using calculations for main-sequence stars with equivalent stellar masses and compare them to calculations of the dynamical ejection timescales from their host GCs.~We suggest that, in general, most BSSs with $v\sin(i)\!>\!70$ km s$^{-1}$ formed no longer than $\sim\!300$ Myr ago in cluster core regions and may be subsequently ejected from their parent GCs. We find two fast rotating BSSs in NGC\,3201 (A4) and in NGC\,6218 (C1) likely in the early process of ejection which must have experienced strong dynamical events no longer than some $\sim$10$^5$ yrs, as well as one rapidly rotating BSS in $\omega$Cen which, as well, appears to be on a hyperbolic trajectory. These strong dynamical events in which BSSs might be kicked out of their parent GCs are likely to be related to their initial formation process.

We investigate the BSS $v\sin(i)$ values as a function of their photometric properties and find that in $\omega$Cen the blue BSS population with colors $(V\!-\!I)_0\la0.25$ mag shows a significantly larger $v\sin(i)$ dispersion than their red counterparts. This remarkable difference between blue and red BSSs is not obvious in the other two GCs due to the smaller BSS samples.~This is the first time that photometric properties can be related to dynamical properties of BSSs, such as their rotational and differential radial velocity.~We have shown that there is a fundamental bimodality in the $v\sin(i)$ distributions of BSSs and a significant difference in the way BSSs populate the CMD of $\omega$Cen according to their internal dynamics.~This, in turn, is likely related to differences in their formation processes and will be discussed in forthcoming papers of this series. 

\acknowledgments
We thank the anonymous referee for comments which improved the presentation of our results. We gratefully acknowledge support from CONICYT through the ALMA-CONICYT Project No.~37070887, FONDECYT Regular Project No.~1121005, FONDAP Center for Astrophysics (15010003), and BASAL Center for Astrophysics and Associated Technologies (PFB-06), as well as support from the HeidelbergCenter in Santiago de Chile and the {\it Deutscher Akademischer Austauschdienst (DAAD)}. This research has made use the Aladin plot tool and the TOPCAT table manipulation software, found in: http://www.starlink.ac.uk/topcat/.

{\it Facilities:} \facility{Las Campanas (IMACS)}, \facility{HST (ACS)}, \facility{La Silla (WFI)}.

\newpage
\begin{turnpage}
\begin{deluxetable*}{lrrrccrccccrc}
\tablecaption{Properties of blue straggler stars in NGC\,3201}
\tablehead{
\colhead{ID} & \colhead{R.A.}   & \colhead{DEC}    & \colhead{$r/r_c$} &\colhead{$I$}  & \colhead{$V\!-\!I$}  & \colhead{$v_{r}$} & \colhead{$\sigma$\tiny{Mgb\ triplet}} & \colhead{$\sigma$\tiny{H$_{\beta}$}} &  \colhead{$\sigma$\tiny{MgII4481}} & \colhead{$\sigma$\tiny{H$_{\gamma}$}}   & \colhead{$v\sin(i)$} & \colhead{Comments} \\
\colhead{} & \colhead{(J2000)}   & \colhead{(J2000)}  & \colhead{ }  & \colhead{(mag)} &
\colhead{(mag)}    & \colhead{(km s$^{-1}$)}   & \colhead{(km s$^{-1}$)}  & \colhead{(km s$^{-1}$)}  & \colhead{(km s$^{-1}$)}  & \colhead{(km s$^{-1}$)}  & \colhead{(km s$^{-1}$)} & \colhead{}
}
\startdata
NGC\,3201-BSS1 & 154.4188021 & $-46.4216510$ & 0.83 & 15.72 & 0.48 & $499.03\pm1.16$ & 27.9 & 30.2 & 22.3 & 40.0 & $27.2\pm0.6$ \\ 
NGC\,3201-BSS2 & 154.4018238 & $-46.4198751$ & 0.35 & 15.50 & 0.57 & $504.46\pm2.64$ & 98.5 & 84.4 & 66.4 & 78.7 & $84.0\pm0.8$ \\ 
NGC\,3201-BSS3 & 154.4389444 & $-46.4182919$ & 1.66 & 16.74 & 0.54 & $516.73\pm2.77$ & 38.6 & 23.9 & 10.7 & 31.6 & $25.0\pm1.6$ \\ 
NGC\,3201-BSS4 & 154.4329169 & $-46.4137565$ & 1.36 & 16.09 & 0.42 & $500.49\pm1.40$ & 20.4 & 39.1 & 34.3 & 58.2 & $28.2\pm1.9$ \\ 
NGC\,3201-BSS5 & 154.4275446 & $-46.4137541$ & 1.11 & 16.65 & 0.48 & $502.43\pm1.65$ & 24.0 & 28.7 & 21.0 & 38.6 & $26.9\pm1.8$ & SX Phe\\ 
NGC\,3201-BSS6 & 154.3699435 & $-46.4114829$ & 1.55 & 16.56 & 0.53 & $497.15\pm1.89$ & 20.7 & 35.1 & 28.5 & 33.7 & $30.8\pm1.0$ \\ 
NGC\,3201-BSS7(A5) & 154.4097267 & $-46.4102398$ & 0.31 & 16.32 & 0.44 & $589.95\pm19.60$ & 121.1 & 139.5 & 153.7 & 126.6 & $135.2\pm3.5$ & anomalous RV \\ 
NGC\,3201-BSS8 & 154.4183676 & $-46.4027641$ & 0.82 & 16.62 & 0.60 & $500.24\pm0.79$ & 19.3 & 28.2 & 25.2 & 32.2 & $20.3\pm1.7$ \\ 
NGC\,3201-BSS9 & 154.3835311 & $-46.3943716$ & 1.24 & 16.50 & 0.44 & $486.05\pm1.06$ & 55.1 & 44.4 & 44.2 & 50.3 & $46.2\pm1.9$ \\ 
NGC\,3201-BSS10 & 154.4030928 & $-46.3843182$ & 1.30 & 16.81 & 0.50 & $505.71\pm0.46$ & -- & 30.0 & 25.0 & 40.8 & $29.5\pm2.3$  & SX Phe\\ 
NGC\,3201-BSS11 & 154.3841700 & $-46.3772401$ & 1.85 & 16.77 & 0.46 & $499.45\pm2.21$ & -- & 31.1 & 22.1 & 21.0 & $28.5\pm2.2$  & SX Phe\\ 
NGC\,3201-BSS12 & 154.4245365 & $-46.4282991$ & 1.22 & 16.80 & 0.60 & $519.08\pm2.80$ & 14.4 & 28.8 & 10.7 & 30.9 & $15.2\pm0.8$ \\ 
NGC\,3201-BSS13 & 154.4006046 & $-46.4278987$ & 0.72 & 16.35 & 0.56 & $500.09\pm1.81$ & 45.9 & 31.7 & 33.5 & 45.6 & $33.0\pm1.6$ & SX Phe \\ 
NGC\,3201-BSS14 & 154.3863888 & $-46.4382670$ & 1.43 & 16.62 & 0.50 & $497.62\pm0.86$ & -- & 41.5 & 35.8 & 44.8 & $37.1\pm1.9$ \\ 
NGC\,3201-BSS15 & 154.3833665 & $-46.4320987$ & 1.29 & 15.77 & 0.71 & $499.67\pm1.08$ & 18.3 & 23.3 & 19.2 & 25.7 & $19.9\pm0.6$ \\ 
NGC\,3201-BSS16 & 154.4151638 & $-46.4224469$ & 0.71 & 16.39 & 0.43 & $538.25\pm2.17$ & 33.5 & 50.6 & 24.5 & 53.5 & $34.6\pm1.6$ \\ 
NGC\,3201-BSS17 & 154.4032964 & $-46.4136807$ & 0.06 & 16.29 & 0.42 & $487.92\pm1.84$ & 82.2 & 95.7 & 67.9 & 71.9 & $77.9\pm4.2$ \\ 
NGC\,3201-BSS18 & 154.3961836 & $-46.4182555$ & 0.43 & 17.06 & 0.53 & $494.53\pm1.42$ & 19.8 & 37.8 & 14.8 & 34.7 & $36.1\pm1.0$ \\ 
NGC\,3201-BSS19 & 154.3668042 & $-46.4150045$ & 1.69 & 17.08 & 0.58 & $498.17\pm2.54$ & 10.3 & 31.4 & 27.0 & 34.8 & $29.4\pm3.7$ \\ 
NGC\,3201-BSS20 & 154.3593971 & $-46.4223781$ & 2.08 & 15.20 & 0.52 & $499.40\pm1.51$ & 54.3 & 51.4 & 33.2 & 49.5 & $44.1\pm2.1$ \\ 
NGC\,3201-BSS21 & 154.4336752 & $-46.4011826$ & 1.49 & 16.88 & 0.56 & $508.64\pm1.64$ & -- & -- & -- & 82.6 & $82.6\pm13.8$ \\ 
NGC\,3201-BSS22 & 154.3765486 & $-46.4000621$ & 1.37 & 16.98 & 0.59 & $498.62\pm2.52$ & 15.1 & 23.7 & 10.3 & 35.1 & $10.5\pm0.2$ \\ 
NGC\,3201-BSS23 & 154.2405779 & $-46.4624675$ & 7.86 & 16.89 & 0.49 & $511.73\pm2.01$ & 23.0 & 25.4 & 10.2 & 27.9 & $10.3\pm0.3$ \\ 
NGC\,3201-BSS24 & 154.3339874 & $-46.4439628$ & 3.52 & 15.84 & 0.37 & $500.08\pm1.04$ & 29.0 & 26.0 & 26.5 & 38.3 & $27.2\pm1.0$  & SX Phe\\ 
NGC\,3201-BSS25 & 154.2510717 & $-46.4092795$ & 7.03 & 16.98 & 0.53 & $500.05\pm1.82$ & 15.3 & 21.2 & 8.1 & 23.9 & $20.2\pm2.2$ \\ 
NGC\,3201-BSS26 & 154.3000518 & $-46.4079181$ & 4.78 & 17.07 & 0.49 & $491.50\pm4.28$ & 28.3 & -- & 11.8 & 35.0 & $24.2\pm3.1$ \\ 
NGC\,3201-BSS27 & 154.2890924 & $-46.4033263$ & 5.29 & 16.73 & 0.44 & $511.07\pm2.57$ & 24.5 & 25.0 & 22.2 & 38.3 & $24.5\pm0.7$ \\ 
NGC\,3201-BSS28 & 154.3138702 & $-46.3868186$ & 4.30 & 16.54 & 0.37 & $501.01\pm2.32$ & -- & 31.4 & 22.0 & -- & $25.4\pm2.9$ \\ 
NGC\,3201-BSS29 & 154.2321831 & $-46.3497612$ & 8.42 & 16.54 & 0.54 & $492.50\pm2.78$ & -- & 28.6 & 11.8 & 36.4 & $22.6\pm0.9$ \\ 
NGC\,3201-BSS30 & 154.3946270 & $-46.4417115$ & 1.41 & 16.69 & 0.50 & $507.86\pm1.90$ & 33.0 & 38.6 & 10.5 & 34.0 & $37.0\pm0.9$ \\ 
NGC\,3201-BSS31 & 154.4625961 & $-46.4328405$ & 2.89 & 16.34 & 0.54 & $511.61\pm1.40$ & 23.5 & 27.8 & 27.5 & 46.8 & $25.7\pm2.2$  & SX Phe\\ 
NGC\,3201-BSS32 & 154.4506853 & $-46.4316515$ & 2.35 & 15.87 & 0.55 & $522.94\pm2.07$ & 16.7 & 28.3 & 19.1 & 40.6 & $19.9\pm0.6$  & SX Phe\\ 
NGC\,3201-BSS33 & 154.3564765 & $-46.4288435$ & 2.29 & 16.58 & 0.52 & $507.21\pm1.09$ & 14.0 & 31.5 & 9.7 & 32.1 & $14.1\pm0.2$ \\ 
NGC\,3201-BSS34 & 154.3606626 & $-46.3985699$ & 2.08 & 16.37 & 0.29 & $469.66\pm0.79$ & 46.3 & 46.4 & 31.4 & 60.0 & $46.9\pm3.1$ \\ 
NGC\,3201-BSS35 & 154.4470979 & $-46.3888138$ & 2.29 & 16.72 & 0.59 & $498.72\pm2.58$ & 20.3 & 35.0 & 17.7 & 33.9 & $32.3\pm1.3$ \\ 
NGC\,3201-BSS36 & 154.4823471 & $-46.3775343$ & 3.98 & 16.02 & 0.41 & $504.25\pm2.15$ & -- & 26.5 & 24.3 & -- & $24.4\pm1.0$ \\ 
NGC\,3201-BSS37 & 154.4337417 & $-46.3495153$ & 3.23 & 16.73 & 0.45 & $503.97\pm1.49$ & -- & 32.2 & 27.2 & 46.0 & $32.0\pm0.8$ 
\enddata
\tablecomments{Variability has been checked in the literature. The variability type is listed accordingly from the \cite{mazur03} catalog.  }\label{tab:n3201}
\end{deluxetable*}

\begin{deluxetable*}{lrrrccrccccrc}
\tablecaption{Properties of blue straggler stars in NGC\,5139}
\tablehead{
\colhead{ID} & \colhead{R.A.}   & \colhead{DEC}    & \colhead{$r/r_c$} &\colhead{$I$}  & \colhead{$V\!-\!I$}  & \colhead{$v_{r}$} & \colhead{$\sigma$\tiny{Mgb\ triplet}} & \colhead{$\sigma$\tiny{H$_{\beta}$}} &  \colhead{$\sigma$\tiny{MgII4481}} & \colhead{$\sigma$\tiny{H$_{\gamma}$}}   & \colhead{$v\sin(i)$} & \colhead{Comments} \\
\colhead{} & \colhead{(J2000)}   & \colhead{(J2000)}  & \colhead{ }  & \colhead{(mag)} &
\colhead{(mag)}    & \colhead{(km s$^{-1}$)}   & \colhead{(km s$^{-1}$)}  & \colhead{(km s$^{-1}$)}  & \colhead{(km s$^{-1}$)}  & \colhead{(km s$^{-1}$)}  & \colhead{(km s$^{-1}$)} & \colhead{}
}
\startdata
NGC\,5139-BSS1 & 201.6671729 & $-47.4507634$ & 1.05 & 16.61 & 0.46 & $209.63\pm2.02$ & -- & 36.4 & 8.0 & 53.6 & $39.0\pm1.0$ \\ 
NGC\,5139-BSS2 & 201.6474051 & $-47.4957164$ & 1.32 & 17.23 & 0.45 & $257.97\pm4.95$ & -- & -- & -- & 46.6 & $46.6\pm60.0$ \\ 
NGC\,5139-BSS3 & 201.6536561 & $-47.4941363$ & 1.15 & 17.16 & 0.45 & $284.99\pm18.59$ & -- & 158.6 & -- & 139.9 & $141.6\pm10.0$ & Ec B\\ 
NGC\,5139-BSS4 & 201.6688540 & $-47.4936837$ & 0.79 & 16.90 & 0.42 & $214.34\pm1.51$ & -- & -- & -- & 45.9 & $45.9\pm15.4$ \\ 
NGC\,5139-BSS5 & 201.6722773 & $-47.4920870$ & 0.70 & 16.33 & 0.43 & $214.86\pm5.09$ & -- & -- & 29.6 & 26.2 & $26.5\pm6.2$ \\ 
NGC\,5139-BSS6 & 201.6829639 & $-47.4890776$ & 0.43 & 16.75 & 0.27 & $213.66\pm2.43$ & -- & 26.2 & -- & 31.0 & $26.2\pm1.9$ \\ 
NGC\,5139-BSS7 & 201.6841193 & $-47.4606848$ & 0.58 & 16.41 & 0.29 & $220.32\pm3.40$ & -- & 105.4 & -- & 89.7 & $100.3\pm12.6$  & Ec W Uma\\ 
NGC\,5139-BSS8 & 201.7074175 & $-47.5095744$ & 0.81 & 16.79 & 0.50 & $160.22\pm2.46$ & -- & 30.5 & -- & 30.6 & $30.5\pm1.2$ \\ 
NGC\,5139-BSS9 & 201.7199519 & $-47.5043337$ & 0.86 & 16.88 & 0.24 & $224.22\pm3.66$ & -- & -- & -- & 135.7 & $135.7\pm11.3$ \\ 
NGC\,5139-BSS10 & 201.7265152 & $-47.4941424$ & 0.84 & 16.48 & 0.22 & $206.72\pm17.92$ & -- & 95.7 & -- & 117.3 & $99.3\pm8.1$ \\ 
NGC\,5139-BSS11 & 201.7317620 & $-47.4654405$ & 0.95 & 16.96 & 0.38 & $258.38\pm0.36$ & -- & 27.4 & -- & 27.7 & $27.6\pm23.1$ \\ 
NGC\,5139-BSS12 & 201.7358275 & $-47.4606257$ & 1.10 & 16.99 & 0.37 & $204.58\pm4.14$ & -- & 57.8 & -- & 41.7 & $57.6\pm2.1$ \\ 
NGC\,5139-BSS13 & 201.6666650 & $-47.4560710$ & 0.97 & 16.82 & 0.30 & $245.44\pm10.95$ & -- & 155.9 & -- & -- & $155.9\pm60.0$ \\ 
NGC\,5139-BSS14 & 201.6580198 & $-47.4849452$ & 0.99 & 16.61 & 0.37 & $209.96\pm2.31$ & 21.4 & 23.5 & -- & 30.1 & $23.9\pm1.5$ \\ 
NGC\,5139-BSS15 & 201.6395133 & $-47.4844789$ & 1.46 & 16.37 & 0.51 & $210.31\pm1.99$ & -- & 27.3 & -- & 34.0 & $27.3\pm0.9$ \\ 
NGC\,5139-BSS16 & 201.6581363 & $-47.4722170$ & 1.00 & 16.46 & 0.33 & $197.56\pm2.77$ & -- & 26.5 & -- & 32.7 & $26.5\pm4.1$ \\ 
NGC\,5139-BSS17 & 201.7247628 & $-47.4990816$ & 0.86 & 16.63 & 0.49 & $220.69\pm1.78$ & -- & 28.0 & -- & 24.9 & $26.5\pm3.0$ \\ 
NGC\,5139-BSS18 & 201.7291359 & $-47.4910398$ & 0.87 & 16.41 & 0.52 & $197.65\pm2.56$ & 6.9 & 40.1 & -- & 36.8 & $36.8\pm1.9$ \\ 
NGC\,5139-BSS19 & 201.7140406 & $-47.4727299$ & 0.47 & 17.02 & 0.36 & $191.38\pm7.33$ & -- & 49.3 & 17.2 & 39.0 & $26.2\pm3.1$ \\ 
NGC\,5139-BSS20 & 201.6674593 & $-47.4155037$ & 1.78 & 16.53 & 0.36 & $212.79\pm1.29$ & -- & -- & -- & 55.9 & $55.9\pm24.7$ & SX Phe\\ 
NGC\,5139-BSS21 & 201.8690796 & $-47.4609422$ & 4.39 & 16.19 & 0.24 & $288.28\pm4.21$ & -- & 83.9 & -- & 118.0 & $86.0\pm11.2$  & Ec W Uma\\ 
NGC\,5139-BSS22 & 201.6148640 & $-47.4542688$ & 2.17 & 16.97 & 0.28 & $232.24\pm3.36$ & -- & 18.8 & -- & 44.0 & $24.9\pm9.8$ \\ 
NGC\,5139-BSS23 & 201.5776652 & $-47.4529202$ & 3.09 & 15.92 & 0.43 & $238.91\pm1.77$ & -- & 42.0 & -- & 41.3 & $42.0\pm1.0$ \\ 
NGC\,5139-BSS24 & 201.5707576 & $-47.5602644$ & 3.79 & 16.22 & 0.41 & $237.57\pm2.59$ & -- & 37.2 & 27.8 & 43.5 & $34.6\pm1.2$ \\ 
NGC\,5139-BSS25 & 201.5286272 & $-47.4466751$ & 4.34 & 17.12 & 0.36 & $234.85\pm6.59$ & -- & 52.7 & -- & -- & $52.7\pm5.3$ \\ 
NGC\,5139-BSS26 & 201.5960187 & $-47.5380422$ & 2.95 & 16.67 & 0.45 & $238.94\pm1.26$ & -- & 33.9 & -- & 27.5 & $33.7\pm1.5$ \\ 
NGC\,5139-BSS27 & 201.6309929 & $-47.5355475$ & 2.19 & 16.91 & 0.41 & $220.16\pm4.75$ & -- & 28.8 & -- & 35.7 & $29.4\pm4.4$ \\ 
NGC\,5139-BSS28 & 201.5851316 & $-47.5333108$ & 3.14 & 16.76 & 0.31 & $232.69\pm1.16$ & -- & 37.5 & -- & 48.2 & $37.7\pm2.0$ & SX Phe\\ 
NGC\,5139-BSS29 & 201.6808083 & $-47.5323744$ & 1.40 & 16.89 & 0.45 & $226.91\pm2.18$ & -- & 32.4 & -- & 29.5 & $32.3\pm1.4$ \\ 
NGC\,5139-BSS30 & 201.6461014 & $-47.5312315$ & 1.83 & 16.01 & 0.42 & $225.47\pm3.26$ & -- & 64.3 & 31.6 & 58.3 & $57.2\pm2.0$ \\ 
NGC\,5139-BSS31 & 201.5374748 & $-47.5300453$ & 4.23 & 17.30 & 0.32 & $247.63\pm2.96$ & -- & 44.7 & -- & 55.2 & $44.8\pm4.2$ & \small{unclass. variable}\\ 
NGC\,5139-BSS32 & 201.5194063 & $-47.4386111$ & 4.61 & 16.42 & 0.47 & $285.25\pm4.83$ & -- & 45.2 & -- & 34.5 & $45.0\pm1.6$ \\ 
NGC\,5139-BSS33 & 201.6595011 & $-47.5269720$ & 1.53 & 17.02 & 0.34 & $231.19\pm2.85$ & -- & 28.3 & -- & 39.3 & $29.8\pm4.5$ & SX Phe\\ 
NGC\,5139-BSS34 & 201.6188360 & $-47.5174123$ & 2.19 & 16.84 & 0.36 & $190.82\pm2.60$ & -- & 35.2 & -- & 42.6 & $35.3\pm3.0$ & SX Phe\\ 
NGC\,5139-BSS35 & 201.5950964 & $-47.5100496$ & 2.69 & 16.69 & 0.25 & $233.78\pm2.96$ & -- & 55.1 & -- & 39.4 & $39.9\pm6.9$ \\ 
NGC\,5139-BSS36 & 201.6628750 & $-47.4828704$ & 0.86 & 16.26 & 0.20 & $241.15\pm4.49$ & -- & 176.2 & -- & 132.3 & $165.6\pm60.0$ \\ 
NGC\,5139-BSS37 & 201.6401835 & $-47.4768963$ & 1.44 & 16.27 & 0.28 & $232.61\pm5.24$ & -- & 164.9 & -- & 134.5 & $163.6\pm8.2$ \\ 
NGC\,5139-BSS38 & 201.5895012 & $-47.4673294$ & 2.73 & 16.62 & 0.16 & $205.38\pm11.20$ & -- & 137.0 & 100.6 & 161.0 & $141.0\pm16.8$ \\ 
NGC\,5139-BSS39 & 201.5372625 & $-47.4193069$ & 4.32 & 16.80 & 0.39 & $188.53\pm2.66$ & -- & -- & -- & 39.4 & $39.4\pm4.6$ \\ 
NGC\,5139-BSS40 & 201.7957272 & $-47.5138816$ & 2.65 & 16.59 & 0.28 & $271.06\pm3.56$ & -- & 59.3 & -- & 68.0 & $62.0\pm12.5$ \\ 
NGC\,5139-BSS41 & 201.8253892 & $-47.4920064$ & 3.27 & 16.58 & 0.28 & $238.78\pm2.08$ & -- & 47.0 & -- & 63.2 & $51.8\pm2.9$ \\ 
NGC\,5139-BSS42 & 201.7917640 & $-47.4869658$ & 2.41 & 16.82 & 0.29 & $251.05\pm4.53$ & -- & 50.4 & 27.6 & 40.8 & $44.2\pm7.3$ \\ 
NGC\,5139-BSS43 & 201.8721714 & $-47.4836168$ & 4.44 & 16.61 & 0.22 & $258.09\pm1.20$ & -- & 42.4 & -- & 40.2 & $42.1\pm10.7$ \\ 
NGC\,5139-BSS44 & 201.7957317 & $-47.4734398$ & 2.51 & 17.16 & 0.37 & $237.16\pm5.34$ & -- & 20.7 & 11.8 & 43.3 & $20.9\pm1.2$ & SX Phe\\ 
NGC\,5139-BSS45 & 201.8421508 & $-47.4644526$ & 3.70 & 16.84 & 0.31 & $213.49\pm1.31$ & -- & 38.2 & -- & 33.8 & $37.3\pm8.5$ & SX Phe\\ 
NGC\,5139-BSS46 & 201.7624194 & $-47.4626258$ & 1.72 & 16.87 & 0.35 & $215.68\pm2.80$ & -- & 26.3 & -- & -- & $26.3\pm3.6$ & SX Phe\\ 
NGC\,5139-BSS47 & 201.6974659 & $-47.4835698$ & 0.10 & 16.87 & 0.32 & $206.61\pm2.05$ & -- & 53.5 & 32.7 & 44.0 & $52.4\pm2.2$ 
\enddata
\tablecomments{Variability has been checked in the literature. The variability type is listed accordingly from the \cite{kaluzny04} catalog.  }\label{tab:n5139}
\end{deluxetable*}

\begin{deluxetable*}{lrrrccrccccrc}
\tablecaption{Properties of blue straggler stars in NGC\,6218}
\tablehead{
\colhead{ID} & \colhead{R.A.}   & \colhead{DEC}    & \colhead{$r/r_c$} &\colhead{$I$}  & \colhead{$V\!-\!I$}  & \colhead{$v_{r}$} & \colhead{$\sigma$\tiny{Mgb\ triplet}} & \colhead{$\sigma$\tiny{H$_{\beta}$}} &  \colhead{$\sigma$\tiny{MgII4481}} & \colhead{$\sigma$\tiny{H$_{\gamma}$}}   & \colhead{$v\sin(i)$} & \colhead{Comments} \\
\colhead{} & \colhead{(J2000)}   & \colhead{(J2000)}  & \colhead{ }  & \colhead{(mag)} &
\colhead{(mag)}    & \colhead{(km s$^{-1}$)}   & \colhead{(km s$^{-1}$)}  & \colhead{(km s$^{-1}$)}  & \colhead{(km s$^{-1}$)}  & \colhead{(km s$^{-1}$)}  & \colhead{(km s$^{-1}$)} & \colhead{}
}
\startdata
NGC\,6218-BSS1 & 251.8045378 & $-1.9641762$ & 1.24 & 16.98 & 0.51 & $-42.32\pm3.46$ & -- & 27.6 & 13.5 & -- & $16.5\pm1.4$ & PM\\ 
NGC\,6218-BSS2 & 251.8012506 & $-1.9617263$ & 1.17 & 16.14 & 0.60 & $-41.03\pm2.67$ & -- & 21.7 & 18.7 & 23.5 & $20.7\pm2.7$ \\ 
NGC\,6218-BSS3 & 251.8220439 & $-1.9616391$ & 1.40 & 16.14 & 0.41 & $-47.41\pm1.98$ & -- & 59.4 & 33.8 & 51.3 & $37.4\pm2.8$ & PM\\ 
NGC\,6218-BSS4 & 251.8340891 & $-1.9525921$ & 1.92 & 16.81 & 0.42 & $-45.63\pm3.52$ & 77.2 & 55.9 & 32.5 & 49.9 & $48.4\pm2.6$ & PM\\ 
NGC\,6218-BSS5 & 251.8132924 & $-1.9523092$ & 0.43 & 16.80 & 0.59 & $-41.06\pm4.83$ & 12.3 & 24.6 & 16.8 & 20.1 & $22.0\pm1.7$ \\ 
NGC\,6218-BSS6 & 251.8145951 & $-1.9514253$ & 0.47 & 16.59 & 0.40 & $-31.14\pm4.15$ & 47.9 & 39.6 & 37.0 & 45.5 & $43.9\pm2.6$ & PM\\ 
NGC\,6218-BSS7(C1) & 251.8063401 & $-1.9505411$ & 0.26 & 16.85 & 0.60 & $47.03\pm4.37$ & 104.3 & 76.7 & -- & -- & $94.9\pm6.0$ & anomalous RV \\ 
NGC\,6218-BSS8 & 251.8387859 & $-1.9501600$ & 2.26 & 17.28 & 0.56 & $-4.04\pm17.93$ & 13.7 & 18.3 & -- & -- & $14.1\pm1.4$ \\ 
NGC\,6218-BSS9 & 251.7870452 & $-1.9468526$ & 1.68 & 17.04 & 0.49 & $-47.33\pm1.50$ & 23.3 & 26.1 & 19.2 & 31.7 & $23.6\pm2.0$ & PM\\ 
NGC\,6218-BSS10 & 251.7970773 & $-1.9465010$ & 0.93 & 16.56 & 0.62 & $-49.20\pm3.81$ & 19.2 & 20.2 & 11.6 & 25.8 & $19.1\pm0.4$ & PM \\ 
NGC\,6218-BSS11 & 251.8237833 & $-1.9421356$ & 1.22 & 17.07 & 0.58 & $-60.67\pm1.93$ & 27.0 & -- & 17.6 & 23.2 & $19.6\pm1.8$ & PM\\ 
NGC\,6218-BSS12 & 251.8201952 & $-1.9419307$ & 0.98 & 16.58 & 0.40 & $-24.66\pm2.43$ & 43.9 & -- & 66.8 & 74.3 & $65.0\pm7.8$ \\ 
NGC\,6218-BSS13 & 251.7993065 & $-1.9386797$ & 1.05 & 16.61 & 0.42 & $-57.21\pm1.68$ & 26.5 & 24.6 & 22.9 & 34.6 & $25.8\pm1.7$ \\ 
NGC\,6218-BSS14 & 251.8120966 & $-1.9255506$ & 1.76 & 16.86 & 0.41 & $-43.41\pm3.38$ & -- & 42.8 & 27.8 & 26.7 & $30.4\pm1.5$ \\ 
NGC\,6218-BSS15 & 251.8028747 & $-1.9212309$ & 2.13 & 17.20 & 0.57 & $-46.08\pm2.18$ & 14.1 & 33.1 & 12.4 & -- & $21.6\pm2.2$ & PM\\ 
NGC\,6218-BSS16 & 251.7576414 & $-2.0245085$ & 6.97 & 16.41 & 0.51 & $-31.07\pm6.48$ & -- & -- & 28.1 & 56.5 & $28.8\pm3.1$ \\ 
NGC\,6218-BSS17 & 251.7720747 & $-2.0175436$ & 5.95 & 17.00 & 0.53 & $-0.05\pm17.34$ & -- & 19.6 & 26.4 & 27.6 & $19.7\pm1.3$ \\ 
NGC\,6218-BSS18 & 251.8362750 & $-1.9887073$ & 3.69 & 17.17 & 0.33 & $-55.31\pm1.40$ & -- & 27.5 & 25.0 & 25.7 & $27.0\pm0.8$ & PM\\ 
NGC\,6218-BSS19 & 251.7635062 & $-1.9793477$ & 4.18 & 16.66 & 0.57 & $-58.38\pm2.93$ & -- & 22.0 & 26.7 & 33.6 & $23.6\pm1.3$ & PM\\ 
NGC\,6218-BSS20 & 251.7952891 & $-1.9720366$ & 2.07 & 16.12 & 0.26 & $-45.48\pm2.46$ & -- & 30.8 & 19.8 & 33.9 & $20.6\pm1.2$ & PM \\ 
NGC\,6218-BSS21 & 251.7883659 & $-1.9716625$ & 2.36 & 17.52 & 0.46 & $-39.58\pm4.21$ & -- & 15.2 & -- & 17.2 & $15.7\pm1.9$ \\ 
NGC\,6218-BSS22 & 251.7620804 & $-1.9604371$ & 3.68 & 16.43 & 0.35 & $-51.02\pm1.31$ & -- & 36.7 & 26.0 & 47.7 & $29.3\pm0.8$ & PM\\ 
NGC\,6218-BSS23 & 251.7697465 & $-1.9577561$ & 3.07 & 15.50 & 0.51 & $-39.51\pm3.90$ & 73.3 & 49.6 & 30.9 & 35.5 & $33.9\pm1.7$ & PM\\ 
NGC\,6218-BSS24 & 251.7846309 & $-1.9550058$ & 1.92 & 17.27 & 0.42 & $-49.02\pm0.91$ & 9.8 & 25.6 & 31.5 & 12.7 & $18.5\pm1.9$ & PM\\ 
NGC\,6218-BSS25 & 251.8578769 & $-1.9517236$ & 3.71 & 17.12 & 0.47 & $-64.62\pm3.32$ & 12.4 & 36.1 & -- & 16.1 & $34.6\pm0.8$ \\ 
NGC\,6218-BSS26 & 251.8314323 & $-1.9496878$ & 1.70 & 16.96 & 0.53 & $-53.54\pm3.56$ & 28.1 & 21.3 & 9.2 & 26.7 & $24.0\pm2.3$ \\ 
NGC\,6218-BSS27 & 251.8483850 & $-1.9475534$ & 2.99 & 16.96 & 0.52 & $-48.39\pm1.28$ & 11.1 & 21.6 & 22.2 & 19.1 & $18.1\pm1.9$ & PM\\ 
NGC\,6218-BSS28 & 251.8177463 & $-1.9382958$ & 1.02 & 17.44 & 0.38 & $-58.69\pm3.46$ & 43.2 & -- & 10.3 & 33.8 & $31.9\pm3.9$ \\ 
NGC\,6218-BSS29 & 251.8252519 & $-1.9376208$ & 1.48 & 16.96 & 0.48 & $-52.06\pm3.75$ & 15.3 & 32.6 & 8.0 & 19.9 & $11.5\pm0.3$ \\ 
NGC\,6218-BSS30 & 251.8470756 & $-1.9279559$ & 3.28 & 16.68 & 0.41 & $-45.76\pm4.37$ & 13.2 & 31.4 & 10.2 & 41.1 & $10.4\pm0.4$ & PM\\ 
NGC\,6218-BSS31 & 251.8159135 & $-1.9269878$ & 1.72 & 17.20 & 0.43 & $-51.76\pm2.10$ & 19.9 & 34.9 & 30.9 & 16.8 & $32.5\pm1.2$ \\ 
NGC\,6218-BSS32 & 251.7448127 & $-1.8918332$ & 6.51 & 17.27 & 0.50 & $-57.36\pm3.76$ & 37.9 & 47.5 & 52.9 & 18.1 & $39.0\pm2.5$ 
\enddata
\tablecomments{Variability has been checked in the literature. We find no variables in our sample. Cluster members confirmed by the \cite{zloczewski12} proper motion catalog are listed as PM.  }\label{tab:n6218}
\end{deluxetable*}
\end{turnpage}

\end{document}